\documentclass{aa}
\usepackage{graphicx}
\usepackage{txfonts}
\begin{document}
   \title{The star formation histories of red and blue low surface brightness disk galaxies}

   \author{G. H. Zhong\inst{1,2,3}\thanks{email: ghzhong@nao.cas.cn},
          Y. C. Liang\inst{1,2}\thanks{email: ycliang@nao.cas.cn},
	  F. Hammer\inst{4}, X. Y. Chen\inst{1,2,3}, L. C. Deng\inst{1,2}, and H. Flores\inst{4}
          }
   \institute{
National Astronomical Observatories,
                Chinese Academy of Sciences, A20 Datun Road, 100012 Beijing, China
         \and
Key Laboratory of Optical Astronomy, National Astronomical
Observatories, Chinese Academy of Sciences, Beijing 100012, China
	 \and
 Graduate School of the Chinese Academy of Sciences, 100049 Beijing, China
         \and
 GEPI, Observatoire de Paris-Meudon, 92195 Meudon, France
             }

   \date{Received ; accepted}

  \abstract
  {}
{We study the star formation histories (SFH) and stellar populations of 213 red and
226 blue nearly face-on low surface brightness disk galaxies (LSBGs), which are selected from the 
main galaxy sample of Sloan Digital Sky Survey (SDSS) Data Release Seven (DR7). 
We also want to compare the stellar populations and SFH between the two groups.
   }
{The sample of both red and blue LSBGs have sufficient signal-to-noise ratio in the
spectral continua. We obtain their absorption-line indices (e.g. $Mg_2$,
$H\delta_A$), $D_n$(4000) and stellar masses from the MPA/JHU catalogs to study their stellar populations
and SFH. Moreover we fit their optical spectra (stellar absorption lines and continua) 
by using the spectral synthesis code STARLIGHT on the basis of the templates of Simple 
Stellar Populations (SSPs).
  }
{We find that red LSBGs tend to be relatively older, higher metallicity, more massive 
and have higher surface mass density than blue LSBGs. The $D_n$(4000)-$H\delta_A$ plane 
shows that perhaps red and blue LSBGs have different SFH: blue LSBGs are more likely 
to be experiencing a sporadic star formation events at the present day, whereas red 
LSBGs are more likely to form stars continuously over the past 1-2 Gyr. Moreover, the 
fraction of galaxies that experienced recent sporadic formation events decreases with 
increasing stellar mass.
Furthermore, two sub-samples are defined for both red and blue LSBGs:
 the sub-sample within the same stellar mass range of 9.5 $\leq$ log$(M_\star/M_\odot)$ 
$\leq$ 10.3,
and the surface brightness limiting sub-sample with $\mu_0(R)$ 
$\geq$ 20.7 mag arcsec$^{-2}$. They show consistent results with the total sample in the corresponding 
relationships, which confirm that our results to compare the blue and red LSBGs are robust.
  }
{}
\authorrunning{G. H. Zhong et al.}
\titlerunning{The star formation histories of red and blue disk LSBGs}

\maketitle


\section{Introduction}

Galaxies with central surface brightness much fainter than the value of $\mu_0(B) = 21.65 \pm 0.3$
mag arcsec$^{-2}$ are well known as low surface brightness galaxies (LSBGs). Although they are faint
compared to the night sky and hard to find, they represent a significant fraction of the number density
of galaxies in the universe (O'Neil \& Bothun 2000; Trachternach et al. 2006) and may comprise up to
half of the local galaxy population (McGaugh et al. 1995).

The most widely studied LSBGs are blue (e.g. Zackrisson et al. 2005; Vorobyov et al. 2009), which 
showed that they appear to have lower metallicity (Burkholder et al. 2001, Haberzettl et al. 2007), 
lower star formation rate (O'Neil et al. 2007), are evolving much more slowly (van den Hoek et al. 2000), 
have larger gas fraction (McGaugh \& de Blok 1997; Schomert et al. 2001), lower galaxy density (Rosehanm 
et al. 2009) and larger amounts of dark matter (de Blok \& McGaugh 1997) than what is typically found 
in normal galaxies.

The wide-field CCD survey of O'Neil et al. (1997) firstly found several red LSBGs with optical
colors compatible with those seen in old stellar populations. In the following, some work compared
the properties of the two groups. Based on the optical-near infrared color-color diagrams of 2 red
and 3 blue LSBGs, Bell et al. (1999) found that red and blue LSBGs have different star formation 
histories (SFH): blue LSBGs are well described by models with low, roughly constant star formation 
rate, whereas red LSBGs are better described by a `faded disk' scenario. Furthermore, with 5 red 
LSBGs, Bell et al. (2000) suggested that the red LSBGs cataloged by O'Neil et al. (1997) are a heterogeneous 
group, which seem to have relatively few common traits.

Due to the considerable uncertainty regarding the SFH of LSBGs, further studies on the properties
of red and blue LSBGs, especially their spectroscopic properties, are important for understanding
their formation and evolution. Moreover, the comparison of their SFH presents a good opportunity
for understanding the global properties of low surface brightness systems.

Nevertheless, the previous studies of red and blue LSBGs have been traditionally carried out with very 
small samples. With the advent of the large sky survey of the Sloan Digital Sky Survey (SDSS), it is now 
possible to extend such studies dramatically in size. Moreover, this enormous amount of high-quality 
data will be undoubtedly important to allow us to study the photometric and spectroscopic properties 
of those galaxies more carefully. We have selected a large sample of LSBGs (Zhong et al. 2008) from 
SDSS-DR4 (Adelman-McCarthy et al. 2006), which consists of much more red and blue LSBGs than before. 
In this work, we select red and blue 
LSBGs from the latest data release of SDSS, the Data Release Seven (DR7)\footnote{http://www.sdss.org/DR7} 
(Abazjian et al. 2009), which greatly extends the sample.
This large sample of galaxies 
will be helpful to explore and compares the SFH of red and blue LSBGs through 
their photometric properties and spectral features, such as the relations of 
$Mg_2$ vs. log$M_*$, $D_n$(4000) vs. $H\delta_A$, 
$D_n$(4000) vs. log$M_*$, surface density and spectral synthesis etc.

This paper is organized as follows. In Section \ref{sec.2}, we describe the selection of the sample.
In Section \ref{sec.3}, we study the SFH of red and blue LSBGs from some property parameters, including
stellar absorption indices, $D_n$(4000), stellar mass, and surface mass density. The stellar populations, 
studied by using spectral synthesis through the STARLIGHT\footnote{http://www.starlight.ufsc.br} code and 
the simple stellar populations (SSPs) of Bruzual \& Charlot (2003), 
are given in Section \ref{sec.4}. The properties of two sub-samples
(surface brightness limiting ($\mu_0(R)$) and stellar mass limiting) are given in Section \ref{sec.5}. 
Then we discuss our results in Section \ref{sec.6},
and summarize this work in Section \ref{sec.7}.

Throughout the paper, a cosmological model with $H_0$ = 70 km s$^{-1}$ Mpc$^{-1}$, $\Omega_M$ = 0.3
and $\Omega_\lambda$ = 0.7 is adopted. All the magnitudes, in Petrosian magnitudes, and colors 
presented here 
are corrected for Galactic extinction and $K$-correction by using the reddening maps of Schlegel et al. 
(1998) and the code provided by Blanton et al. (2003), respectively.

\section{The Sample}
\label{sec.2}

The SDSS is the most ambitious astronomical survey ever undertaken in imaging and spectroscopy
(Stoughton et al. 2002) for hundreds of thousands galaxies. The imaging data are done in drift
scan mode and are 95\% complete for point sources at 22.0, 22.2, 22.2, 21.3, and 20.5 in five
bands ($u$, $g$, $r$, $i$, $z$), respectively. The spectroscopic data provide spectral flux and
wavelength calibrated with 4096 pixels from 3800 to 9200 \AA~at resolution R $\sim$ 1800. Our
sample was selected from the Main Galaxy Sample (MGS) of SDSS-DR7.
Following Zhong et al. (2008), we firstly selected 21,664 nearly face-on disk LSBGs 
from the SDSS-DR7 MGS. 
Then red and blue LSBGs are selected from their $g-r$ vs. $r-i$ diagram.
These color-selected LSBGs are further matched with the spectral catalog to select
those having higher signal-to-noise (S/N) ratio on the spectral continua.
The detailed selection criteria are given below.

\begin{figure*}
\begin{center}
\includegraphics [width=7.2cm, height=7.0cm] {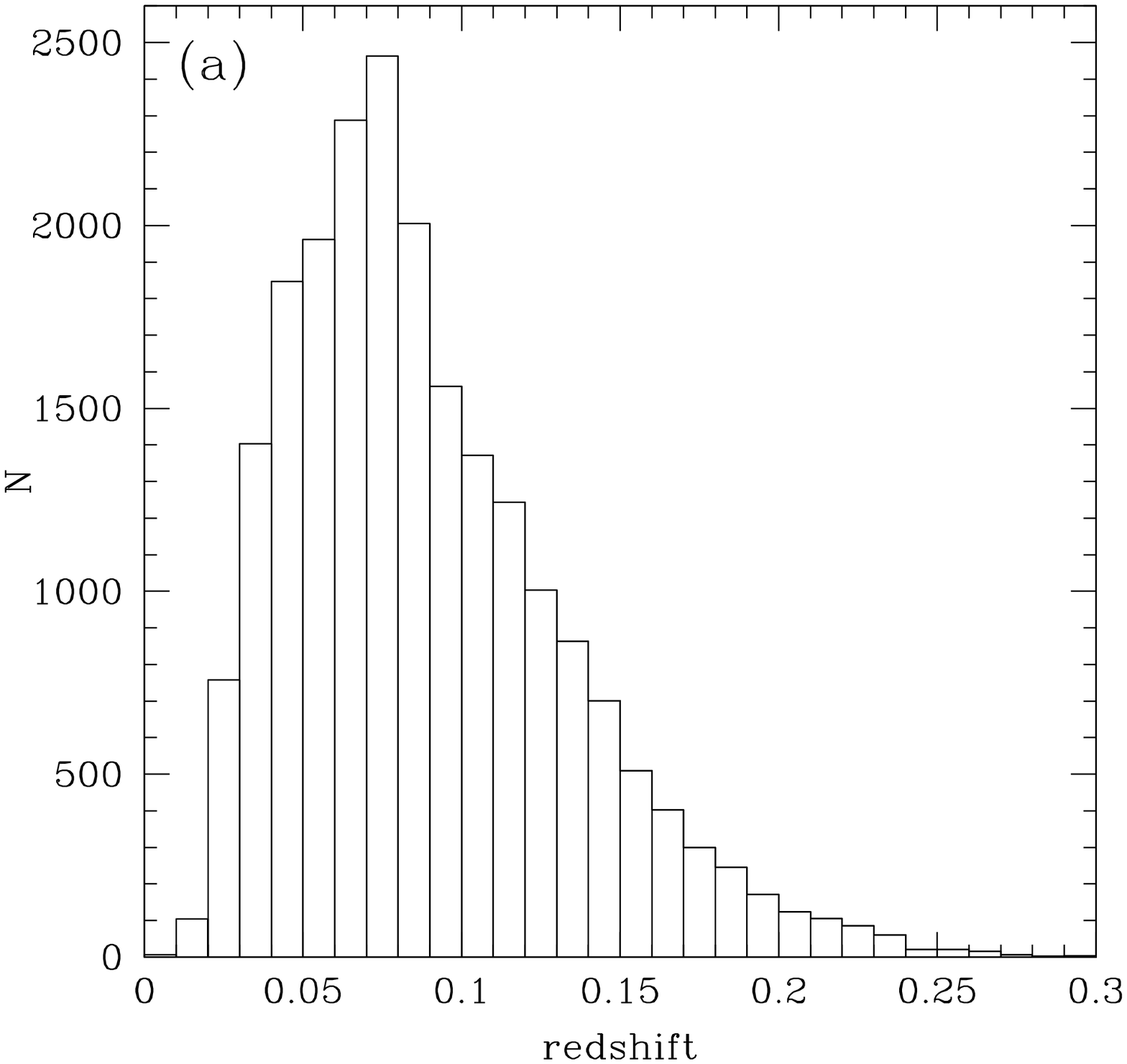}
\includegraphics [width=7.2cm, height=7.0cm] {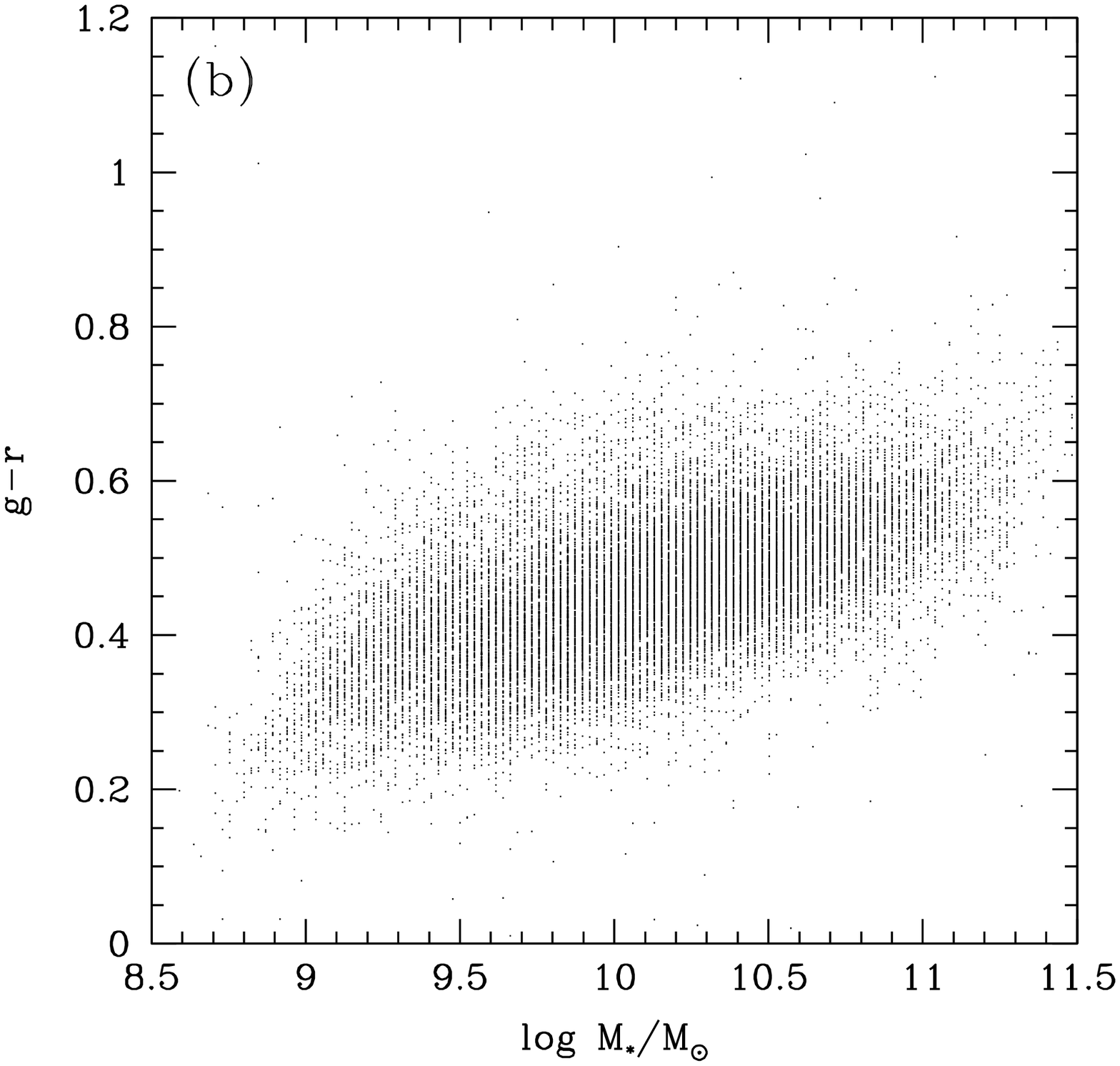}
\includegraphics [width=7.2cm, height=7.0cm] {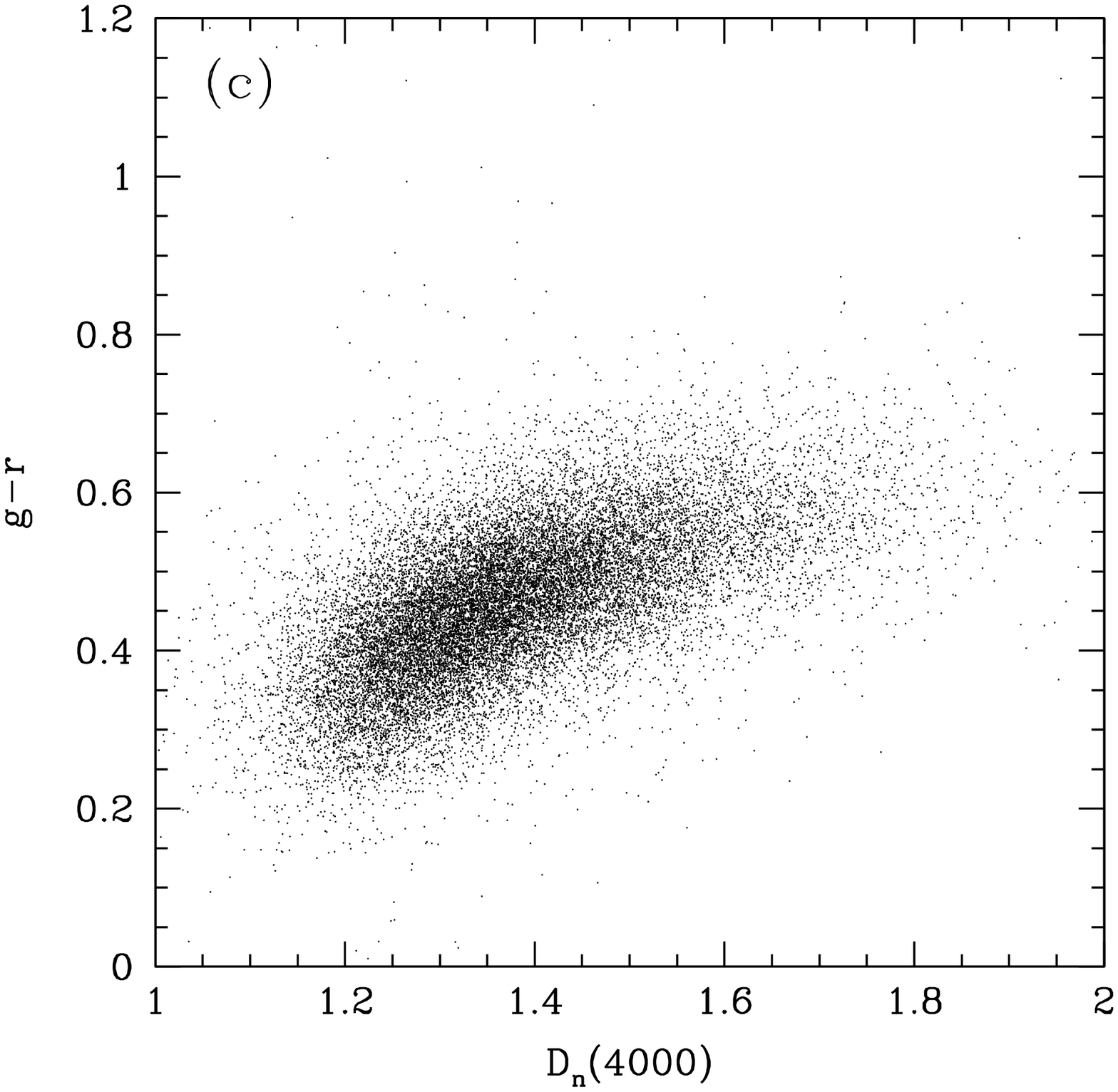}
\includegraphics [width=7.2cm, height=7.0cm] {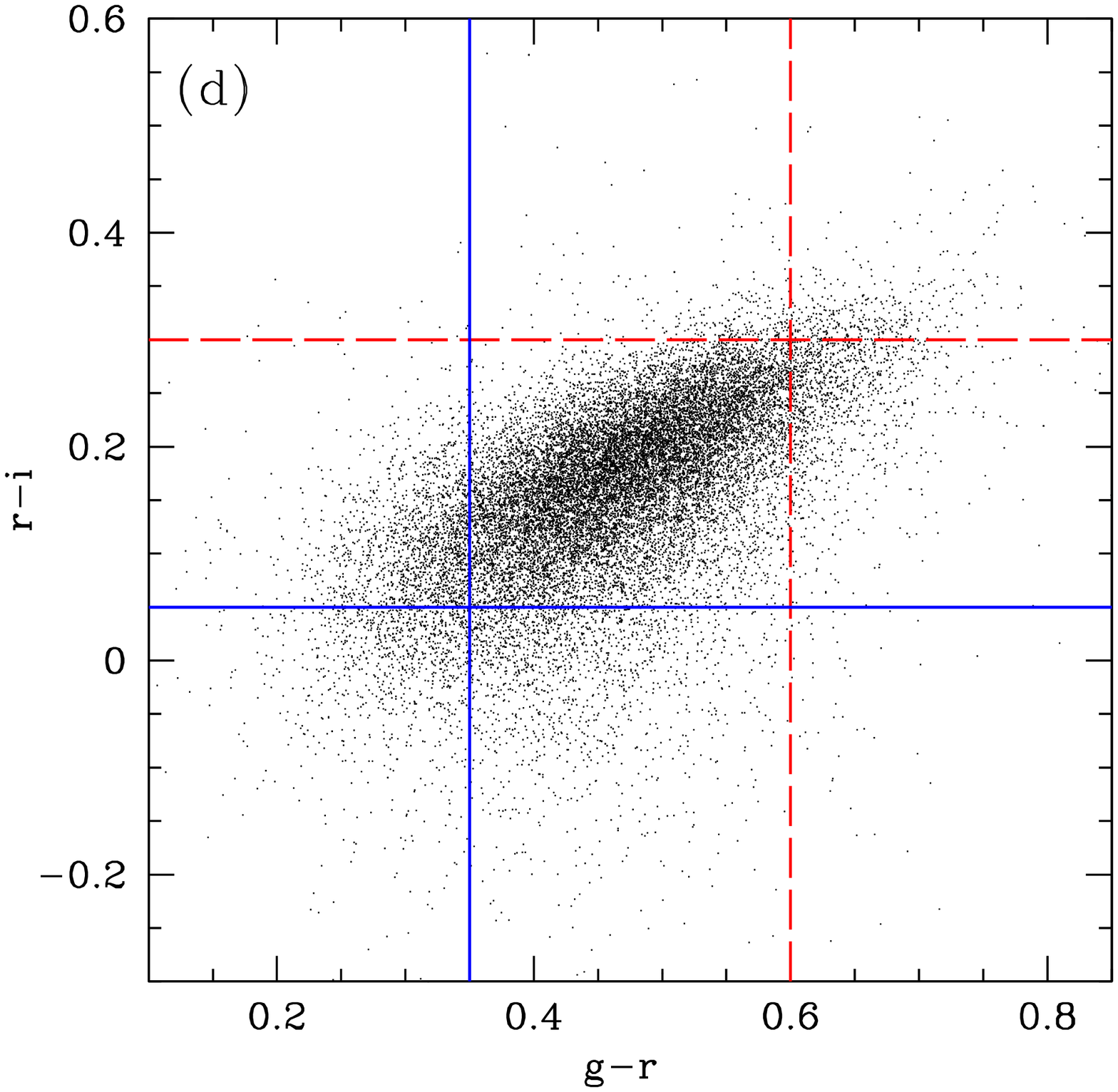}
\caption{The properties of the 21,664 nearly face-on disk LSBGs: (a). histogram distribution of 
redshift; (b). distribution of $g-r$ colors with stellar mass; (c). $g-r$ colors is plotted as a 
function of $D_n$(4000); (d). $g-r$ versus $r-i$ diagram, where the solid and dashed lines are used to define 
the red (top-right corner) and blue (bottom-left corner) LSBGs, respectively (see text).}
\label{fig.0}
\end{center}
\end{figure*}

\begin{figure}
\begin{center}
\includegraphics [width=7.5cm, height=7.2cm] {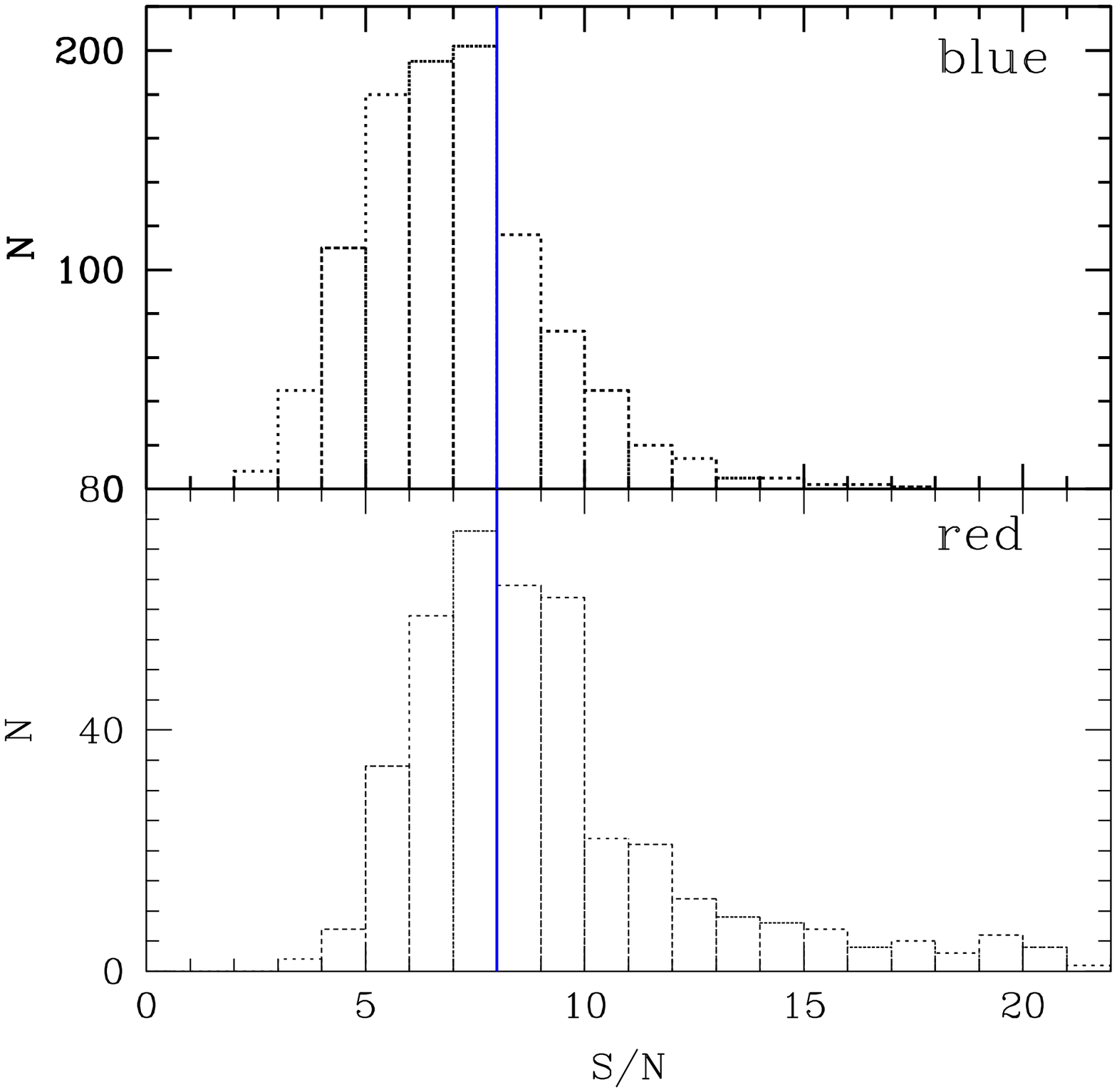}
\caption{The histogram distribution of S/N for the 405 red (bottom) and 1,025 blue (top) LSBGs, respectively.
The median values are 8.5 and 6.8, respectively. The solid lines refer to S/N $=$ 8.0.}
\label{fig.A}
\end{center}
\end{figure}

\begin{enumerate}
\item $fracDev_r <$ 0.25, indicating the fraction of luminosity contributed by the 
de Vaucouleurs profile relative to exponential 
profile in the $r$-band is much smaller (Bernardi et al. 2005; Chang et al. 2006; Shao et al. 2007); 
$b/a >$ 0.75, this corresponds to 
the inclination $i <$ 41.41 degree, which is to select the nearly face-on disk galaxies (Liu et al. 2009)
($a$ and 
$b$ are the semi-major and semi-minor axes of the fitted exponential disk, respectively); 
$M_B < -18.0$, this is to exclude the few dwarf galaxies
($M_B$ is the absolute magnitude in the $B$-band);
$\mu_0(B) \ge$ 22.0 mag arcsec$^{-2}$, this is to select the LSBGs 
(O'Neil et al. 1997; Impey et al. 2001; Boissier et al.  2003). 
After applying the above four selection criteria, 21,664 nearly face-on disk LSBGs are selected. 
Their $\mu_0(B)$ 
are from 22.0 to 24.5 mag arcsec$^{-2}$ with a median value of 22.43 mag arcsec$^{-2}$. 

\item In Fig \ref{fig.0}a, 
we present the histogram distribution of the redshift of this large sample of galaxies,
showing 0.01$<z<$0.27 with the median value of 0.08. 
In Fig. \ref{fig.0}b, we show the distributions of $g-r$ color 
with stellar mass (taken from the MPA/JHU stellar mass catalog as given in Kauffmann et al. 2003a
and Gallazzi et al. 2005). 
It can be seen that there exists a slightly correlation showing that more 
massive LSBGs have redder $g-r$ colors generally. 
In Fig. \ref{fig.0}c, $g-r$ color is plotted as a 
function of $D_n$(4000) (taken from the MPA/JHU spectroscopic catalog as given in Kauffmann et al. 2003a). 
It shows a slightly correlation between $g-r$ color and $D_n$(4000) as well, showing that $g-r$ 
color becomes redder along with the value of $D_n$(4000) going higher. 
Fig. \ref{fig.0}d shows the 
$g-r$ versus $r-i$ diagram for the selected LSBGs. 

Then two categories are selected from Fig. \ref{fig.0}d : 
the blue LSBGs with $g-r < 0.35, r-i < 0.05$ (the bottom left corner 
of the solid lines and the red LSBGs with $g-r > 0.6, r-i > 0.3$ (the top right corner 
of the dashed lines). This step results in 405 red and 1,025 blue LSBGs.

\item Matching with the MPA/JHU\footnote{http://www.mpa-garching.mpg.de/SDSS/DR7} spectroscopic catalog,
404 red and 1,022 blue LSBGs have spectral observations and S/N measurements. 
For doing spectral synthesis studies on the galaxies through fitting their spectral continua and absorption lines,
good S/N are needed for their spectra. Therefore, we only select the galaxies having
median S/N per pixel of the whole spectrum greater than 8.0.
Then 226 red and 276 blue LSBGs are selected. 

Fig. \ref{fig.A} shows the histogram distribution 
of S/N for all the 404 red (bottom) and 1,022 blue (top) LSBGs (with median values of 8.5 and 6.8, respectively),
and the selected sample with S/N$>$ 8.0 (226 red and 276 blue LSBGs) marked by the solid vertical lines (the right parts of the lines).
Furthermore, we remove 13 red and 10 blue LSBGs, since
their spectra are not continuous throughout the whole spectra due to some problems,
which are not suitable to be synthesized.

Finally, we selected 213 red and 266 blue disk LSBGs to study their
SFH and spectral synthesis as presented in Sect.~\ref{sec.3} and Sect.~\ref{sec.4}, respectively.
We call this sample the ``T-sample" (T means total).

\item Furthermore, to minimize/check the effect of B-band surface brightness for the red LSBGs,
we then select a sub-sample by further considering the surface brightness limit $\mu_0(R)$ $\geq$ 20.7 mag 
arcsec$^{-2}$, then 100 red and 262 blue LSBGs are selected. 
This does show that the $B$-band surface brightness selection may benefit to 
select blue LSBGs. However, we hope this will not have much affect on the basic results 
on the properties of these two types of galaxies with blue or red colors.
We call this sub-sample the ``$\mu$-sample", and will study their properties in 
Sect.~\ref{sec.5}.1.

\item For more reliable comparison between red and blue LSBGs with similar stellar mass, 
we further select a sub-sample with 
stellar mass limit of 9.5 $\leq$ log$(M_\star/M_\odot)$ $\leq$ 10.3 from the total sample, 
and then 83 red and 120 blue LSBGs are 
selected to be further studied. We call this sub-sample as ``M-sample".
Their properties will be specially presented in Sect.~\ref{sec.5}.2.
\end{enumerate}

\section{Star Formation History}
\label{sec.3}

In this section, we study the SFH of red and blue LSBGs with some property parameters, such as
$Mg_2$, $H\delta_A$ (Worthey \& Ottaviani 1997), $D_n$(4000) (Balogh et al. 1999), stellar
mass, and surface mass density, all the values of which are taken from or calculated on the basis of
the MPA/JHU catalog.

\begin{figure*}
\begin{center}
\includegraphics [width=7.2cm, height=7.0cm] {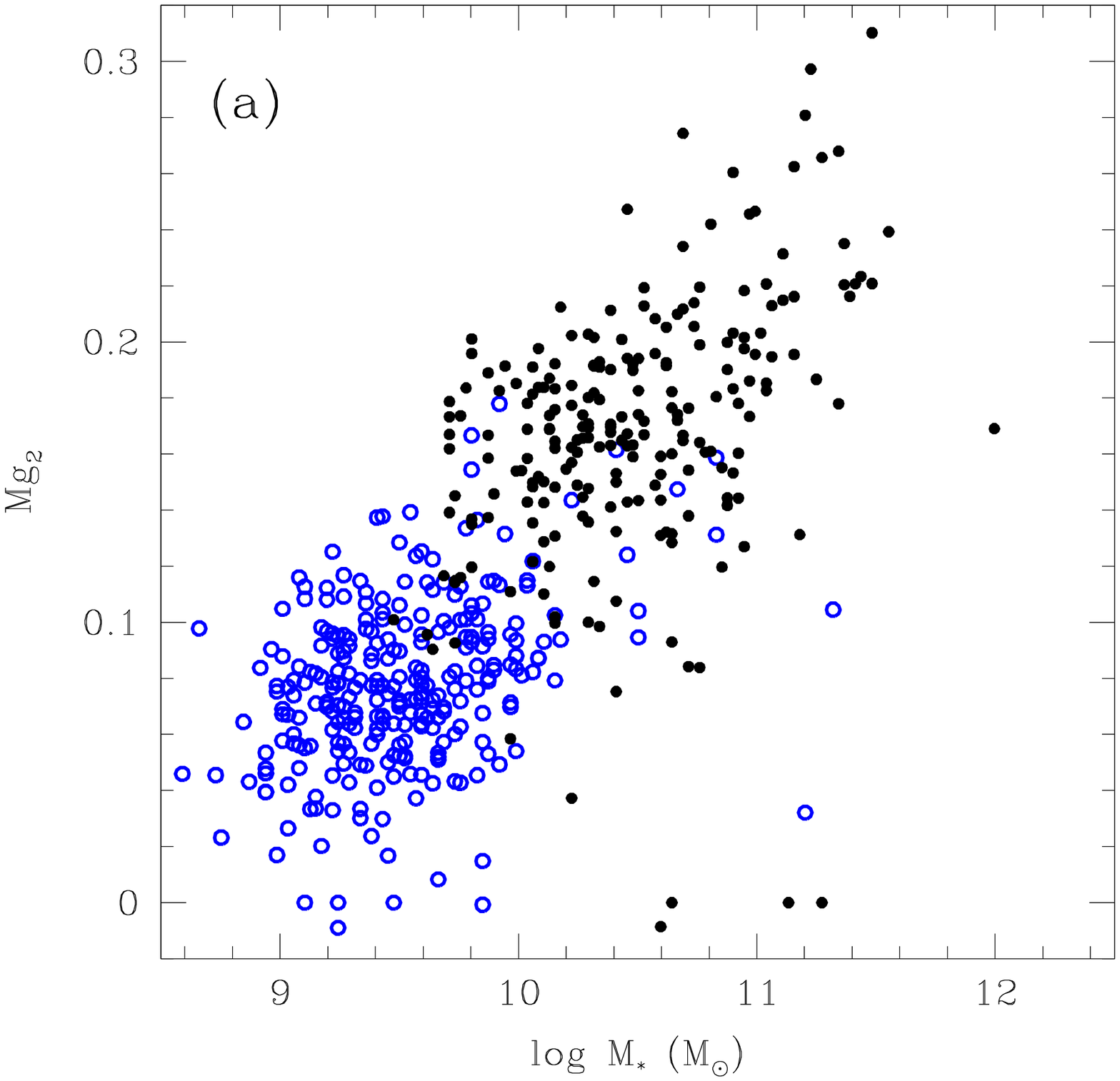}
\includegraphics [width=7.2cm, height=7.0cm] {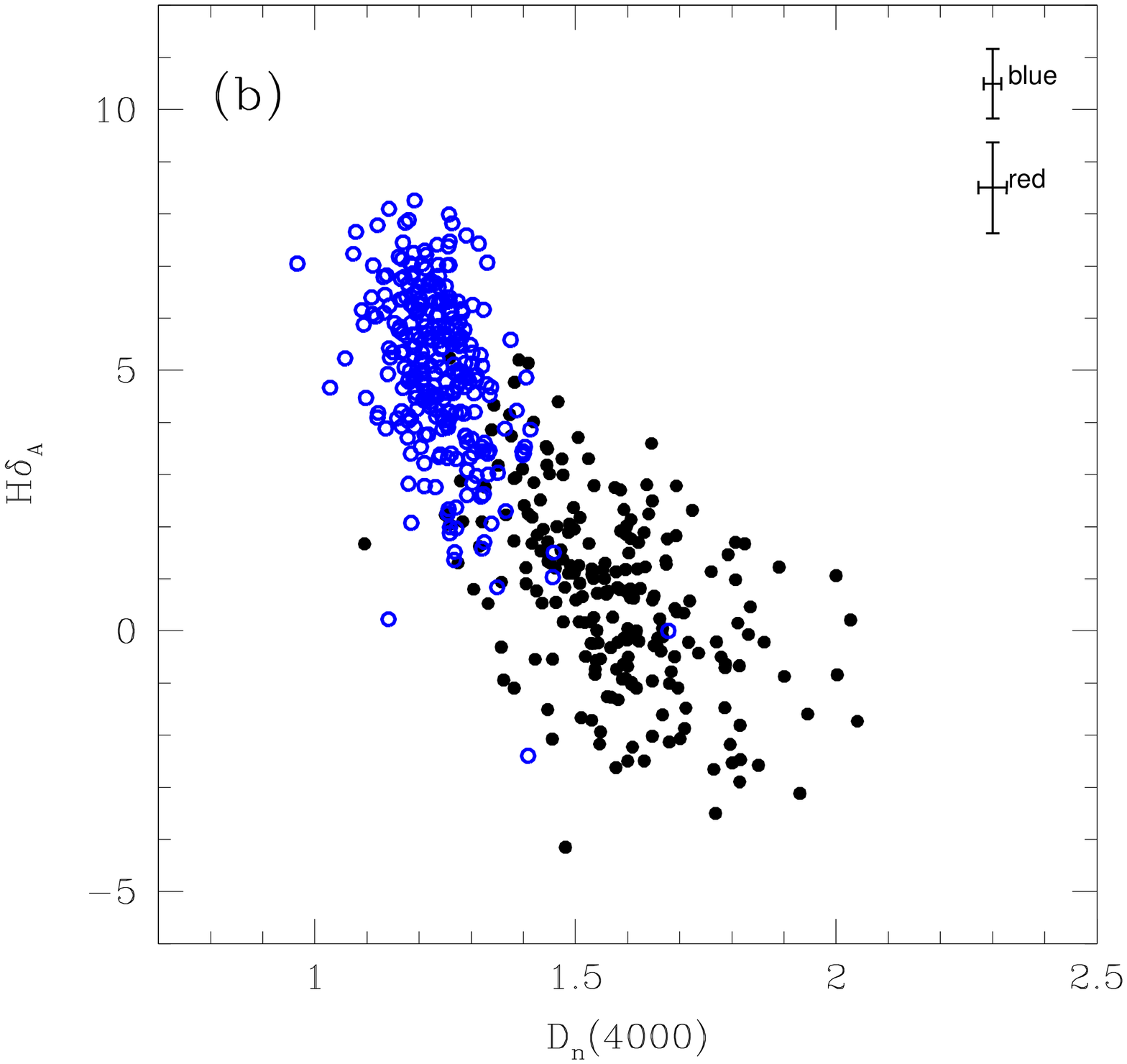}
\includegraphics [width=7.2cm, height=7.0cm] {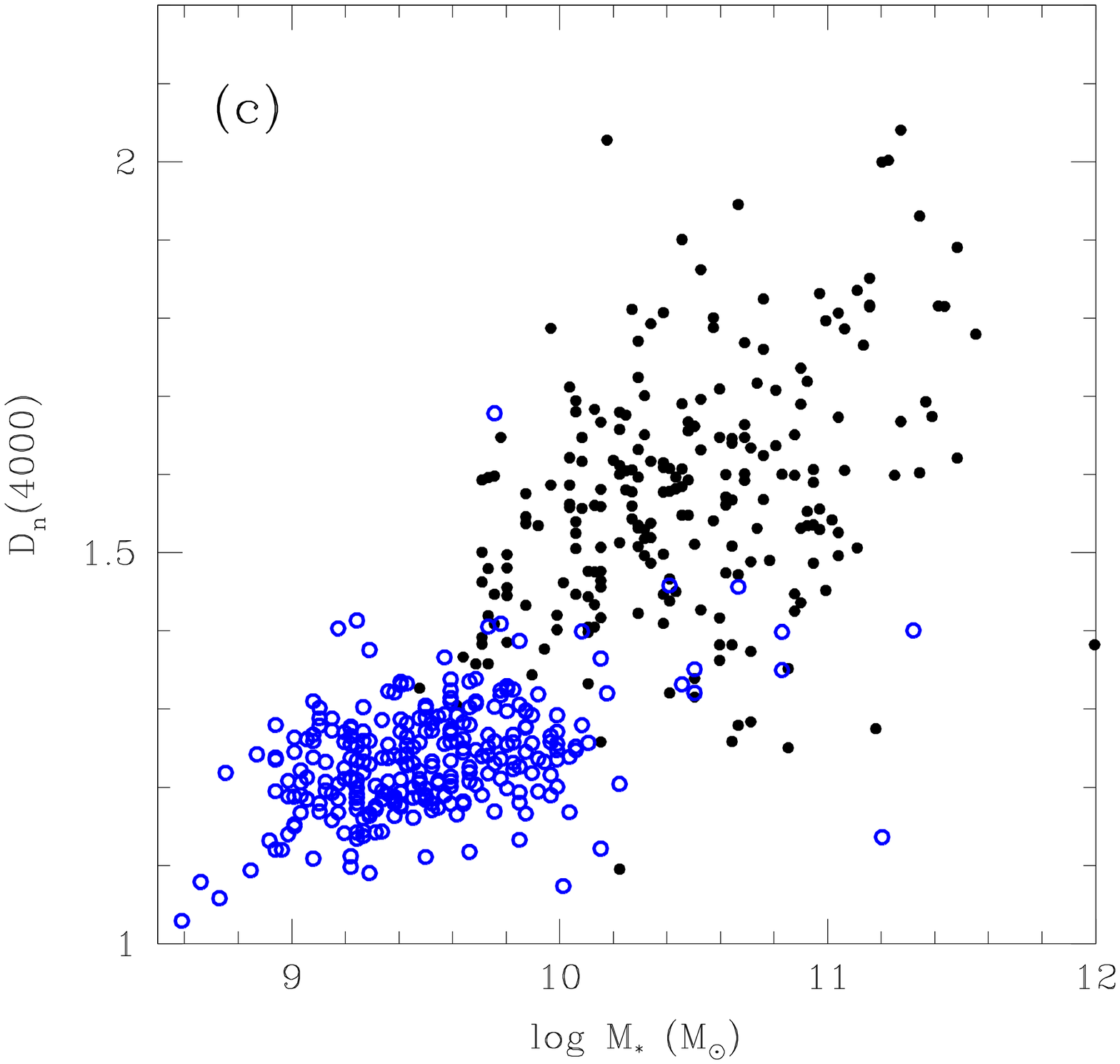}
\includegraphics [width=7.2cm, height=7.0cm] {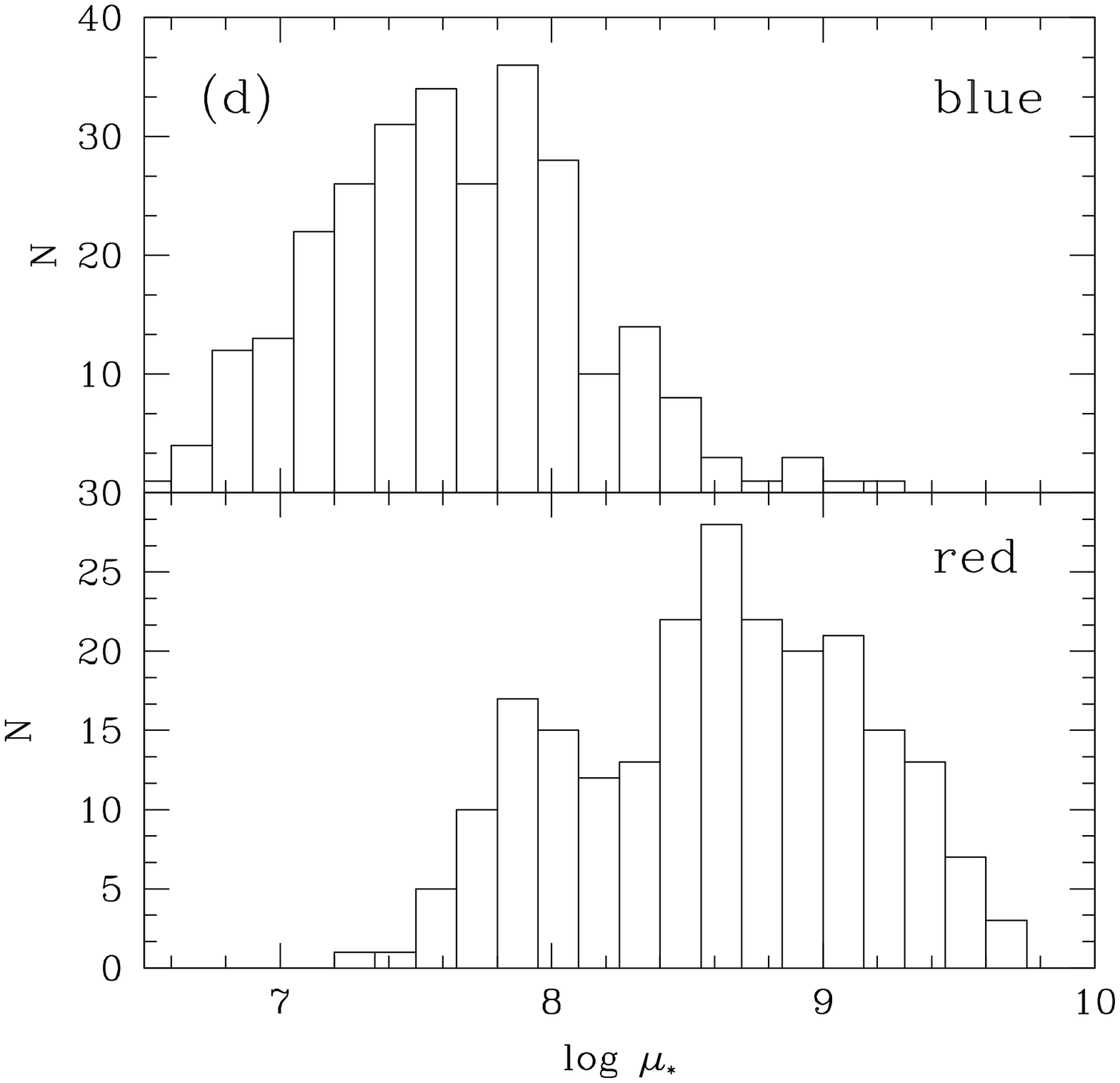}
\caption{The properties of red and blue LSBGs: (a). the distribution of $Mg_2$ as a function of stellar 
mass; (b). $H\delta_A$ is plotted as a function of $D_n$(4000); (c). the relation between $D_n$(4000) 
and stellar mass; (d). histogram distribution of surface mass densities for red (bottom) and blue (top) 
LSBGs. The filled and open circles denote red and blue LSBGs, respectively.}
\label{fig.1}
\end{center}
\end{figure*}

$Mg_2$, containing both $Mg~b$ and $MgH$ absorption, is sensitive to metallicity and responds
very similarly like changes of $\alpha/Fe$ to $Mg~b$ (Thomas et al. 2003). $Mg_2$ increases with
increasing $\alpha/Fe$ ratio. In Fig. \ref{fig.1}a, the distributions of $Mg_2$ with stellar
mass are shown. It suggests that red LSBGs are more massive than the blue ones
generally with the median values of 4.1 $\times$ 10$^{10}$ $M_\odot$ and 7.6 $\times$ 10$^9$ $M_\odot$, 
respectively. 
It is in accordance with the luminosity-metallicity relations that the 
redder colors correspond to galaxies with larger stellar mass (Galaz et al. 2002), 
meaning that more massive galaxies contain more metals and/or older stellar populations 
than lower mass galaxies although with scatters.

It has been discussed by Kauffmann et al. (2003a) that $D_n$(4000)-$H\delta_A$ plane is a powerful 
diagnostic for the SFH of galaxies, e.g. whether galaxies have been forming stars continuously or in 
bursts over the past 1-2Gyr. Therefore, we use the narrow definition of $D_n$(4000) and $H\delta_A$ 
as a diagnostics of the SFH of red and blue LSBGs. In Fig. \ref{fig.1}b, we show the $H\delta_A$ absorption 
index as a function of $D_n$(4000). One can see clearly that nearly all blue LSBGs have lower $D_n$(4000) 
values ($\leq$ 1.4, characteristic of stellar populations with mean ages of less than a few Gyr (Kauffmann 
et al. 2003b)) and stronger $H\delta_A$ absorption than red LSBGs. It means that blue LSBGs have a 
higher fraction of young stars, hence have smaller mass-to-light ratios (in $z$ band, Kauffmann et al. 
2003a), and are more likely to be experiencing a sporadic star formation events at the present day. 
Whereas red LSBGs are distributed in the continual star formation regions suggested by Kauffmann et al. 
(2003a, see their Fig. 6), which means that red LSBGs are more likely to form stars continually over 
the past years. This may be due to their higher stellar mass and corresponding to higher surface mass
density (Kauffmann et al. 2003b). Moreover, the $H\delta_A$ of blue LSBGs drops more rapidly than red 
LSBGs with the increasing $D_n$(4000). 

In Fig. \ref{fig.1}c, $D_n$(4000) is plotted as a function of stellar mass. Considering Fig. \ref{fig.1}b 
and Fig. \ref{fig.1}c together, we notice that the fraction of galaxies that have experienced recent 
sporadic star formation events decreases with increasing stellar mass, which is consistent with Kauffmann 
et al. (2003b, see their Fig. 3).

In Fig. \ref{fig.1}d, we show the histogram distribution of surface mass densities for red
(bottom) and blue (top) LSBGs. Following Kauffmann et al. (2003b), we define the surface mass
density $\mu_\star$ as $0.5M_\star/[\pi R_{50}^2(z)]$, where $R_{50}(z)$ is the Petrosian half-light
radius in the $z$ band. The surface mass density of red LSBGs is higher than blue LSBGs, with
median values 4.0 $\times$ 10$^8$ and 4.0 $\times$ 10$^7$ for red and blue LSBGs, respectively.

It is suggested that galaxies with larger stellar masses tend to have higher surface mass
density (Kauffmann et al. 2003b), but smaller gas mass fractions (Galaz et al. 2002). Thus
the older stellar populations and higher surface mass density of red LSBGs indicate an epoch
of more vigorous star formation over the past years, which has also been suggested by Bell et
al. (1999). The younger stellar populations and lower mean metallicities of blue LSBGs principally indicate
the galaxies are being slow to convert gas into stars (Bell et al. 2000; Schombert
et al. 2001) and are relatively unevolved.

\section{Spectral Synthesis}
\label{sec.4}
\begin{figure*}
\begin{center}
\includegraphics [width=7.5cm, height=7.0cm] {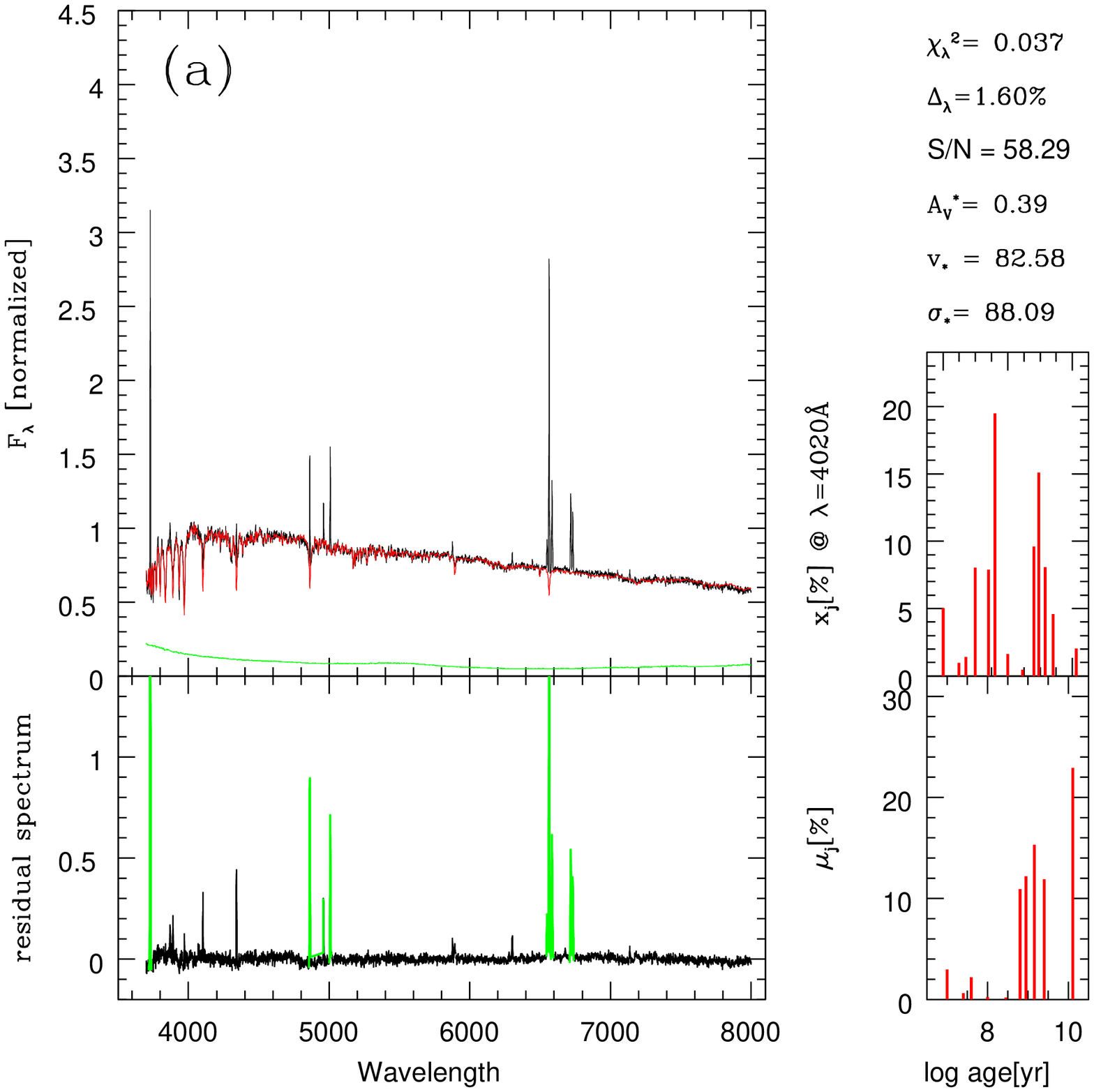}
\includegraphics [width=7.5cm, height=7.0cm] {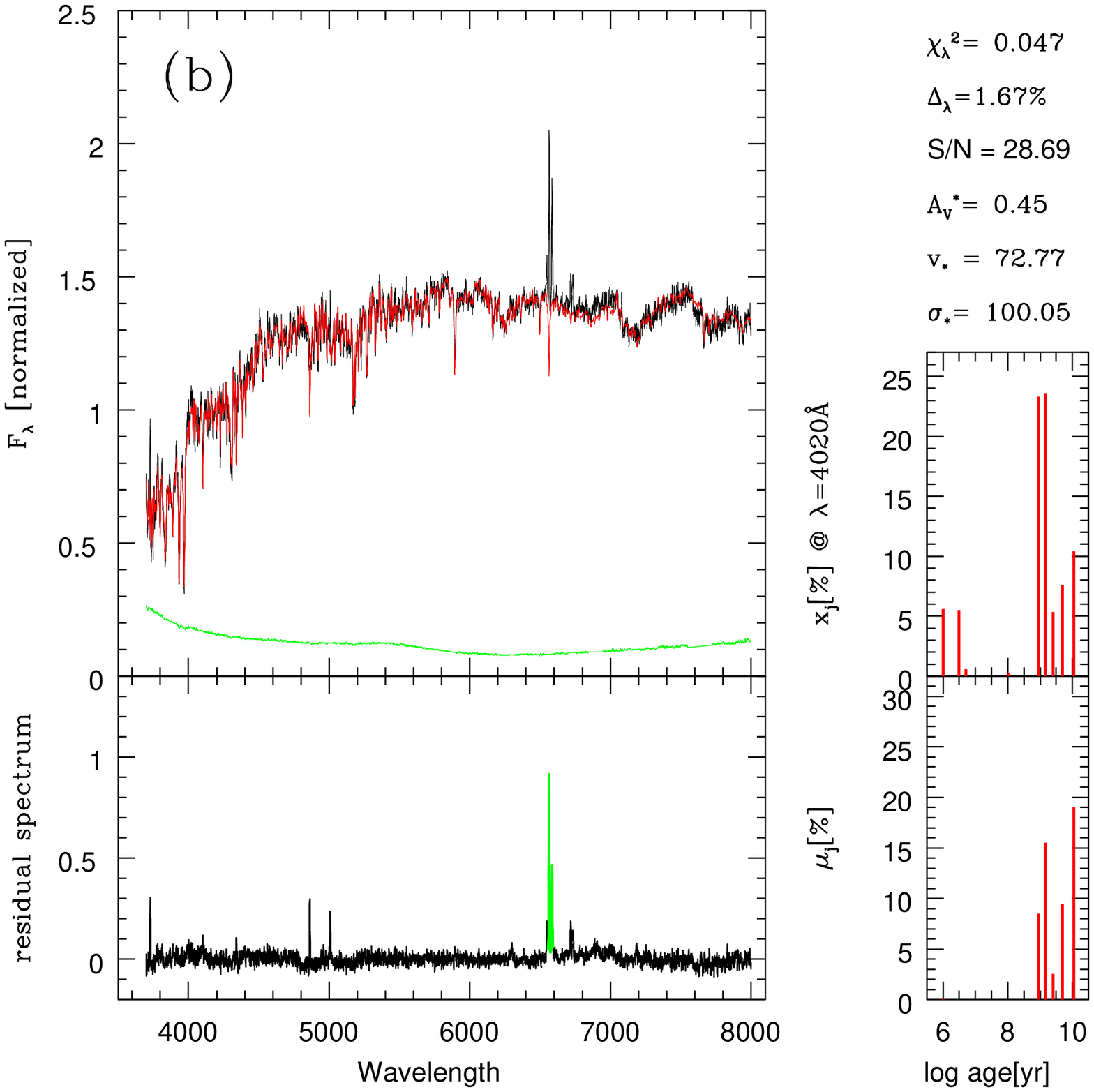}
\caption{The combined spectral synthesis results of blue (left) and red (right) LSBGs by using
STARLIGHT with 45 SSPs from Bruzual \& Charlot (2003). In each of the panel , top left: the synthesis
spectrum (red line), the observed spectrum (black line), and the error spectrum (green line);
bottom left: the residual spectrum, the green lines represent masked regions given by the SDSS flag;
right: the contribution fractions of light (top) and mass (bottom) as a function of the 15 ages
of SSPs. We list six parameters in the top right corners. (Please see on-line color version for
more details.)}
\label{fig.2}
\end{center}
\end{figure*}
\begin{table*}
\caption{Stellar populations of red and blue disk LSBGs. Results of fitting the combined spectra by
using STARLIGHT with 45 SSPs from Bruzual \& Charlot (2003). The contributed light fractions in 3 age-bins (young populations with
age $\leq$ 5 $\times$ 10$^8$ yr, intermediate populations with 6.4 $\times$ 10$^8$ yr $\leq$ age
$\leq$ 2.5 $\times$ 10$^9$ yr and old populations with age $\geq$ 5.0 $\times$ 10$^9$ yr) and 3
metallicities (0.2, 1.0 and 2.5 $Z_\odot$) are presented. It is noticed that the percent fractions
here are a bit different from the $x_j$ component output by STARLIGHT directly.}
\centering
\begin{tabular}{c|c|c|c}
\hline \multicolumn{2}{c|}{SSP}
  & \multicolumn{1}{c|}{blue LSBGs}  &
\multicolumn{1}{c}{red LSBGs}   \\
\hline
age($t$, Gyr) & young ($t$ $\leq$ 0.5) & 58.8 & 38.9  \\
&intermediate (0.64 $\leq$ $t$ $\leq$ 2.5) & 35.3 &  27.8 \\
&old ($t$ $\geq$ 5.0) &  5.9 &   33.3 \\
\hline
$Z/Z_{\odot}$ &0.2 & 58.8 & 35.3\\
 &1 &  29.4 &  47.1 \\
 &2.5 &  11.8 &  17.6 \\
\hline
\end{tabular}
\label{table1}
\end{table*}

Spectral synthesis provides a new way to retrieve information of stellar populations of galaxies
from observational spectra, which is crucial for a deeper understanding of galaxy formation and
evolution. Galaxy spectra contain information about the age and metallicity distributions of
the stars, which in turn reflect the star formation and chemical enrichment histories of the galaxies.

We fit the optical spectra of red and blue LSBGs by using the spectra synthesis code STARLIGHT (Cid
Fernandes et al. 2005; Mateus et al. 2006; Asari et al. 2007). The method consists of fitting an observed 
spectrum $O_\lambda$ with a model $M_\lambda$ combination of $N_*$ spectral components SSPs taken 
from Bruzual \& Charlot (2003). In this work, we take 45 SSPs, including 15 different ages from 1 
Myr to 13 Gyr (i.e. 1, 3, 5, 10, 25, 40, 100, 280, 640, 900 Myr and 1.4, 2.5, 5, 11, 13 Gyr) and 3 
metallicities (i.e. 0.2, 1, and 2.5 $Z_\odot$), the stellar evolutionary tracks of Padova 1994 (Alongi et a.l. 
1993; Girardi et al. 1996), the Initial Mass Function (IMF) of Chabrier (2003), and the extinction law 
of Cardelli et al. (1989) with $R_V = $ 3.1. The Galactic extinctions are corrected by the reddening 
map of Schlegel et al. (1998), then shifted to the rest frame. The range of the spectra is from 3700 
to 8000 \AA~ in step of 1 \AA~ and normalized by the median flux in the 4010 to 4060 \AA~ region
by the median value. During spectral synthesis fitting, we exclude the emission lines, sky lines and 
another four windows (5870-5905 \AA, 6845-6945 \AA,7550-7725 \AA,7165-7210 \AA) as done in Chen et al.
(2009, 2010).

Fig.~\ref{fig.2} shows the spectral fitting results on the combined spectra for blue (left) and red 
(right) LSBGs. There are four sub-panels in each panel: the top left shows the synthesis spectrum 
(red line), the observed spectrum (black line), and the error spectrum (green line); bottom left shows 
the residual spectrum, the green lines represent mask regions given by SDSS flag; right panel shows 
the contributed fractions to light (top) and mass (bottom) from the 15 SSPs with different ages. We 
list the produced six parameters in the top right corners, such as ${\chi_\lambda}^2$, i.e. the 
reduced ${\chi}^2$; the mean relative difference between synthesis and observed spectra $\Delta_\lambda$; 
the S/N in the region of 4730-4780 \AA; $V$-band extinction; the velocity $v_\star$ and the velocity 
dispersion $\sigma_\star$. The contributed light fractions of the stellar populations in age-bin and metallicity-bin 
are presented in Table~\ref{table1}. The three age bins are young populations with age $\leq$ 5 $\times$ 
10$^8$ $yr$, intermediate populations with 6.4 $\times$ 10$^8$ $yr$ $\leq$ age $\leq$ 2.5 $\times$ 10$^9$ 
$yr$ and old populations with age $\geq$ 5.0 $\times$ 10$^9$ $yr$ and their metallicities are 0.2, 1.0 
and 2.5 $Z_\odot$, respectively.

Fig.~\ref{fig.2} and Table~\ref{table1} show that red LSBGs are older than blue LSBGs. Blue
LSBGs are dominated by the young (58.8\%) and intermediate age populations (35.3\%) with a
small fraction of old age populations (5.9\%). Red LSBGs have a very significant fraction of
old age populations (33.3\%, about 27\% larger than blue ones), although they also have
significant young age populations (38.9\%), which suggests that there was an epoch of more
vigorous star formation in red LSBGs in the past. This is consistent with Bell et al. (1999),
who commented that red LSBGs have higher mean stellar ages than blue LSBGs. 

It also shows that the metallicities of red LSBGs are higher than blue LSBGs, however, the 
difference of metallicities between red and blue LSBGs is not as obvious as the difference 
of age populations. The dominate metallicities are $Z_\odot$ for red (47.1\%) and 0.2$Z_\odot$ 
for blue (58.8\%) LSBGs. The fraction of metallicities of 2.5$Z_\odot$ in red LSBGs are 17.6\%, 
which is 6\% higher than blue LSBGs (11.8\%). 
It should be noticed that the dominate contributions on stellar mass 
are all old in both of the two groups. For the uncertainties of these results, we will specially 
discuss them in Section \ref{sec.6}, however, we believe that they are insignificant.

\section{The Sub-samples}
\label{sec.5}
In this section, we select two sub-samples, surface brightness limiting sub-sample ($\mu$-sample) 
and mass limiting sub-samples (M-sample), to discuss the selection effects on surface brightness 
and stellar mass.

\subsection{Surface Brightness Limiting Sub-sample}

\begin{figure*}
\begin{center}
\includegraphics [width=7.2cm, height=7.0cm] {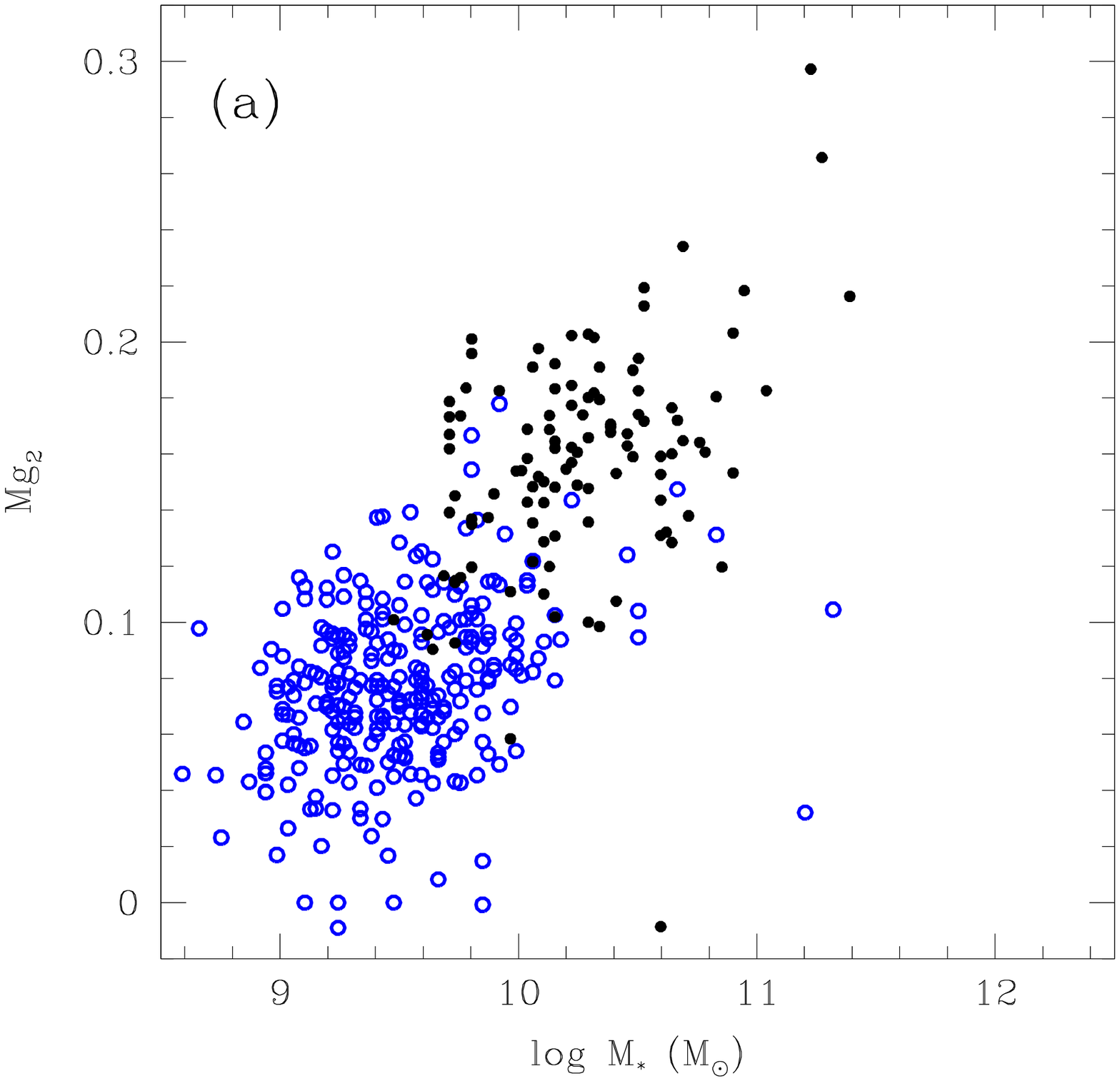}
\includegraphics [width=7.2cm, height=7.0cm] {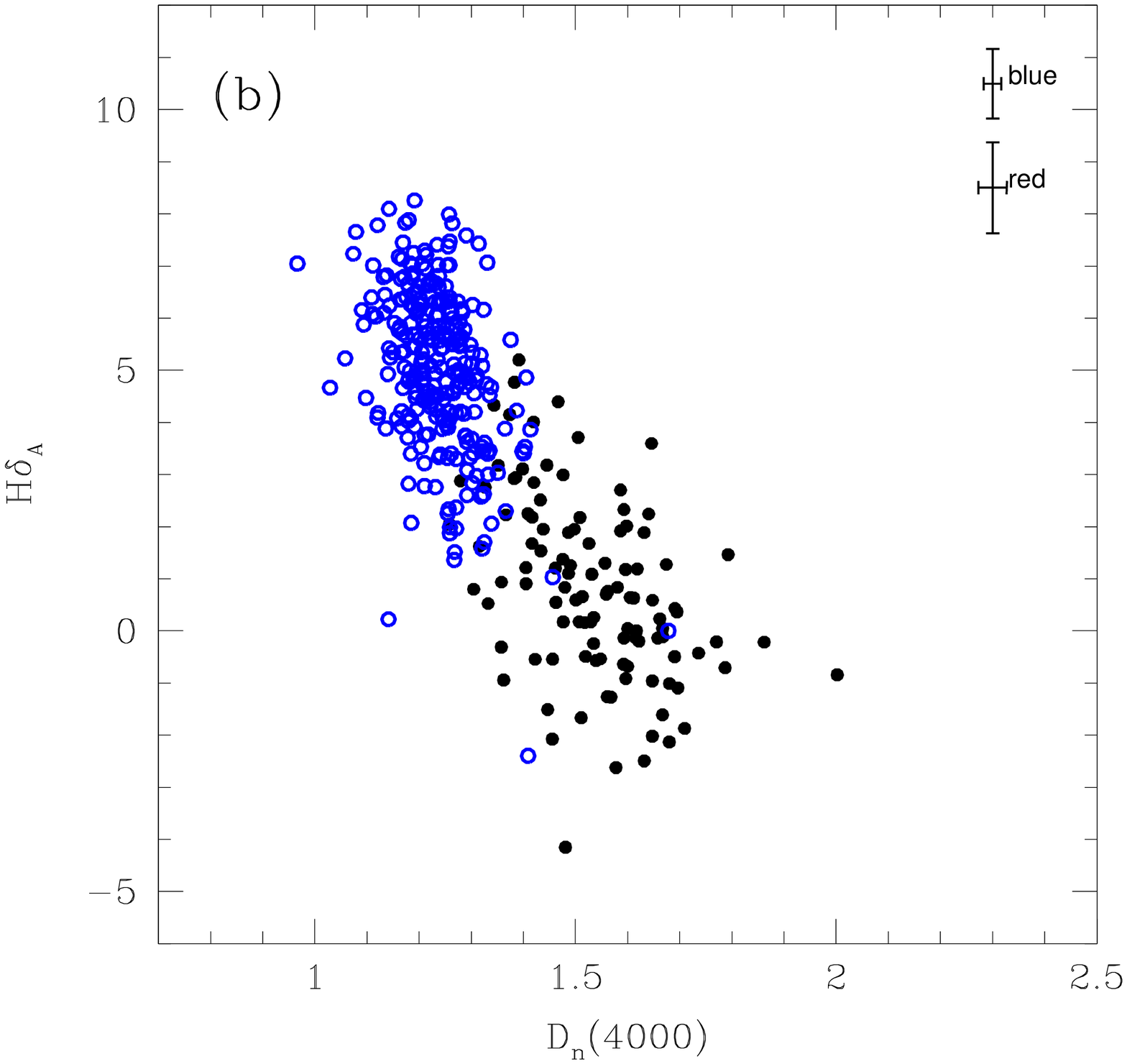}
\includegraphics [width=7.2cm, height=7.0cm] {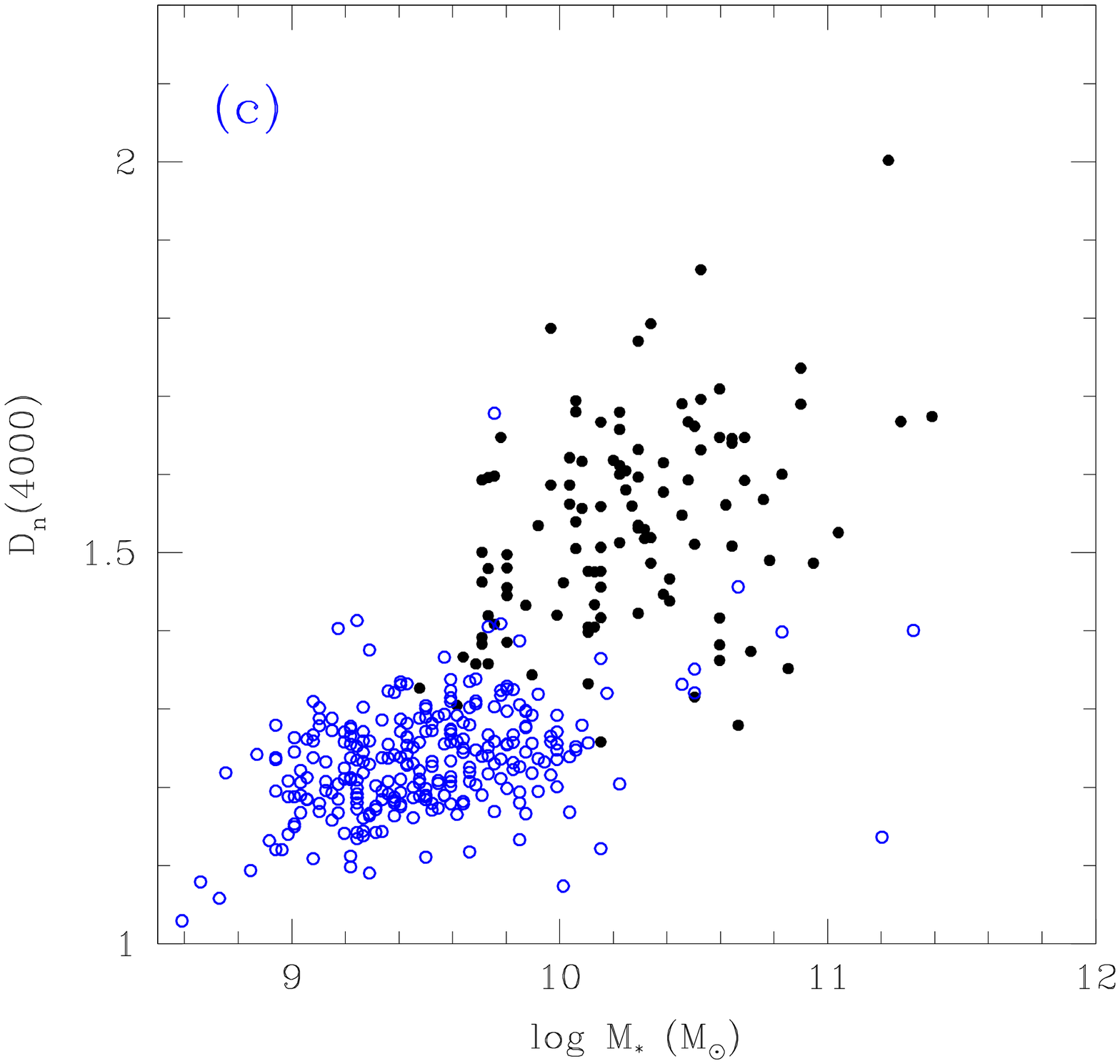}
\includegraphics [width=7.2cm, height=7.0cm] {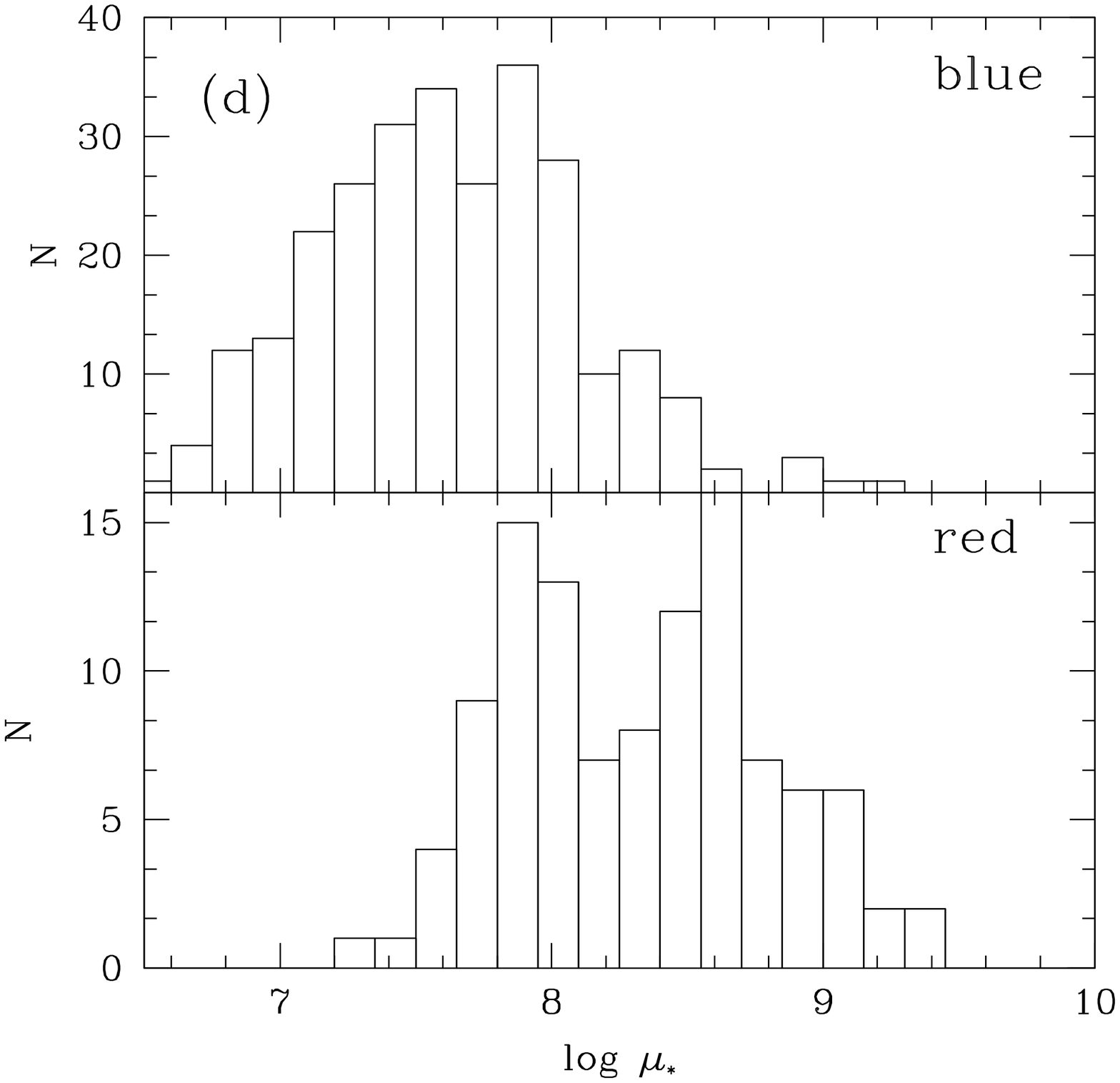}
\caption{The same as Fig.~\ref{fig.1} but for the surface brightness limiting sub-samples further with 
$\mu_0(R)$ $\geq$ 20.7 mag arcsec$^{-2}$.}
\label{fig.3}
\end{center}
\end{figure*}
\begin{table*}
\caption{Stellar populations of red and blue LSBGs sub-samples for 
surface brightness limiting with $\mu_0(R)$ $\geq$ 
20.7 mag arcsec$^{-2}$ and stellar mass limiting with 9.5 $\leq$ log$(M_\star/M_\odot)$ $\leq$ 10.3.}
\centering
\begin{tabular}{c|c|c|c|c|c}
\hline \multicolumn{2}{c|}{SSPs} &
\multicolumn{2}{c|}{surface brightness limiting sub-samples} &
\multicolumn{2}{c}{stellar mass limiting sub-samples}
\\

\hline 
& & \multicolumn{1}{c}{blue LSBGs}  &
\multicolumn{1}{c|}{red LSBGs}  &
\multicolumn{1}{c|}{blue LSBGs} & \multicolumn{1}{c}{red LSBGs} \\
\hline
age($t$, Gyr) & young ($t$ $\leq$ 0.5) & 61.9 & 38.9 & 60.0 & 41.7 \\
&intermediate (0.64 $\leq$ $t$ $\leq$ 2.5)  & 33.3 & 38.9 & 35.0 & 25.0 \\
&old ($t$ $\geq$ 5.0) & 4.8 & 22.2 &  5.0 &   33.3  \\
\hline
$Z/Z_{\odot}$ &0.2 & 42.9 & 44.5 & 60.0 & 41.6 \\
 &1 & 33.3  & 22.2 & 25.0 &  29.2 \\
 &2.5  & 23.8  & 33.3 &  15.0 &  29.2  \\
\hline
\end{tabular}
\label{table2}
\end{table*}

Our selection criterion of LSBGs is $\mu_0(B)$, which is a blue filter surface brightness criterion, thus 
some red LSBGs may not be intrinsically of low surface brightness. Therefore, we further define a double 
criterion to limit the surface brightness in both the blue and red filters by adding the $\mu_0(R)$ selection 
criterion for our total sample. Courteau (1996) found there is a well-defined upper cutoff at $\mu_0(R) =$ 
20.08 $\pm$ 0.55 mag arcsec$^{-2}$ (also see the introduction of Galaz et al. 2002) for LSBGs. We choose 
$\mu_0(R)$ $\geq$ 20.7 mag arcsec$^{-2}$ as our red filter selection criterion and then obtain 100 red and 
262 blue LSBGs as our surface brightness limiting sub-sample, the $\mu$-sample. 
Following Fig. \ref{fig.1} and Table \ref{table1},
their properties are showed in Fig. \ref{fig.3}, 
and the results of spectral synthesis are showed in Table \ref{table2}.

The results show that red LSBGs are affected much by the red filter selection criterion and 113 red LSBGs are 
removed from this. Only 4 blue LSBGs are removed by this criterion. Therefore, the distributions of red and blue 
LSBGs show some differences in the $Mg_2$ vs. log$(M_\star/M_\odot)$ (Fig. \ref{fig.3}a) and $D_n$(4000) vs. log$(M_\star/M_\odot)$
(Fig. \ref{fig.3}c) relations. Comparing Fig. \ref{fig.1}d and Fig. \ref{fig.3}d, we can see that the samples 
having relatively higher surface mass density are removed. 
However the median surface mass density values of red and blue LSBGs are  2.5 $\times$
10$^8$ and 4.0 $\times$ 10$^7$, respectively, which are not so much different from 
the total sample (4.0 $\times$ 10$^8$ and 4.0 $\times$ 10$^7$, respectively).
The distributions of the sub-sample galaxies in the $D_n$(4000)-$H\delta_A$ 
plane (Fig. \ref{fig.3}b) also do not show much difference from the total sample as given in Fig. \ref{fig.1}b. 
Comparing Table \ref{table2} and Table \ref{table1}, 
 it shows that this $\mu_0(R)$ limit nearly does not 
change the fraction of stellar populations of blue LSBGs, since only 4 blue LSBGs are removed. 
For red LSBGs, 
the $\mu_0(R)$ limit increases the intermediate stellar population by 11\% and correspondingly decreases 
the old stellar population 
by 11\%, and the young stellar population nearly has no changes.
These mean that this $\mu_0(R)$ limit removes some red LSBGs with older stellar populations.

\subsection{Mass Limiting Sub-sample}
\begin{figure*}
\begin{center}
\includegraphics [width=7.5cm, height=7.0cm] {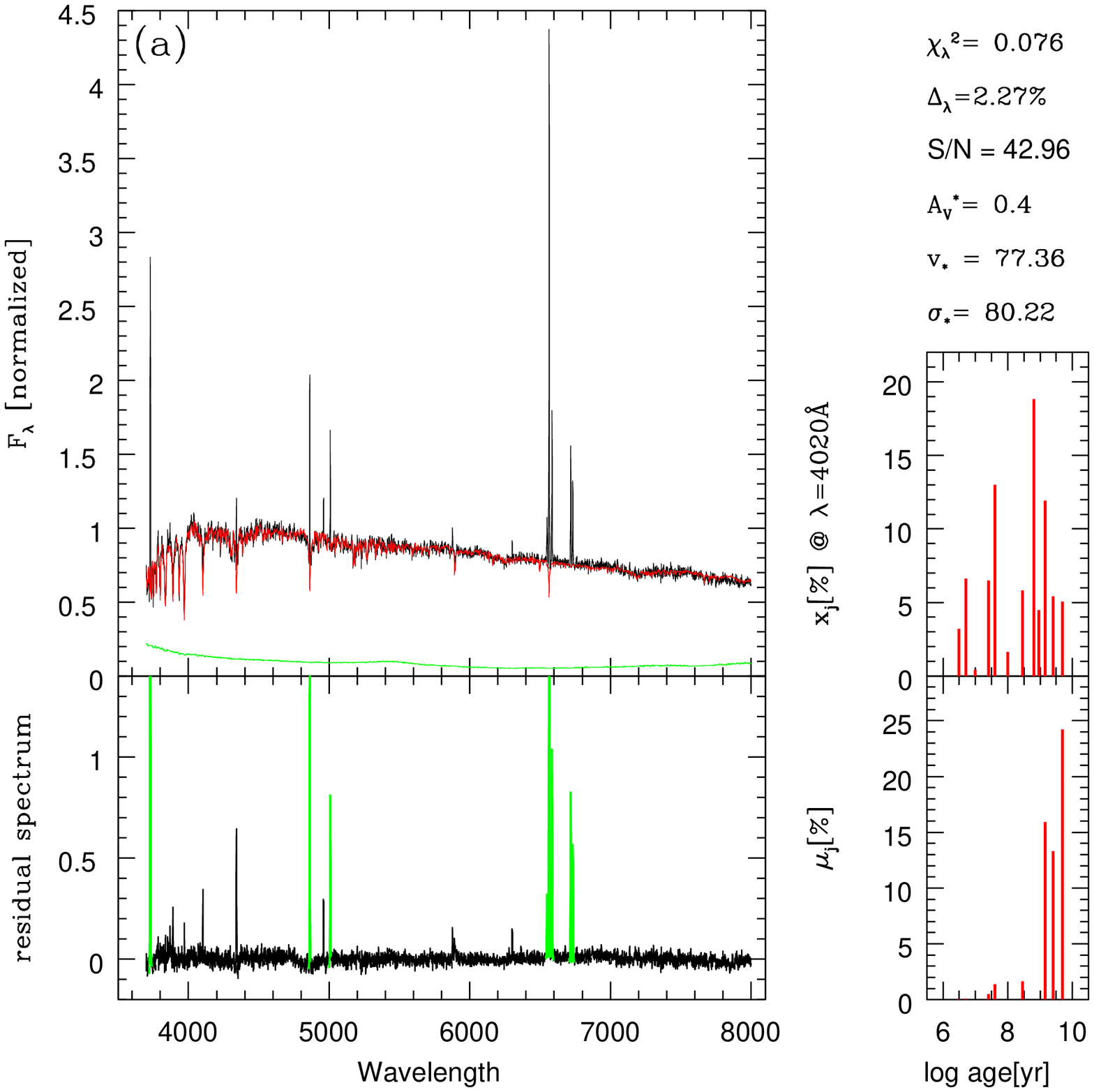}
\includegraphics [width=7.5cm, height=7.0cm] {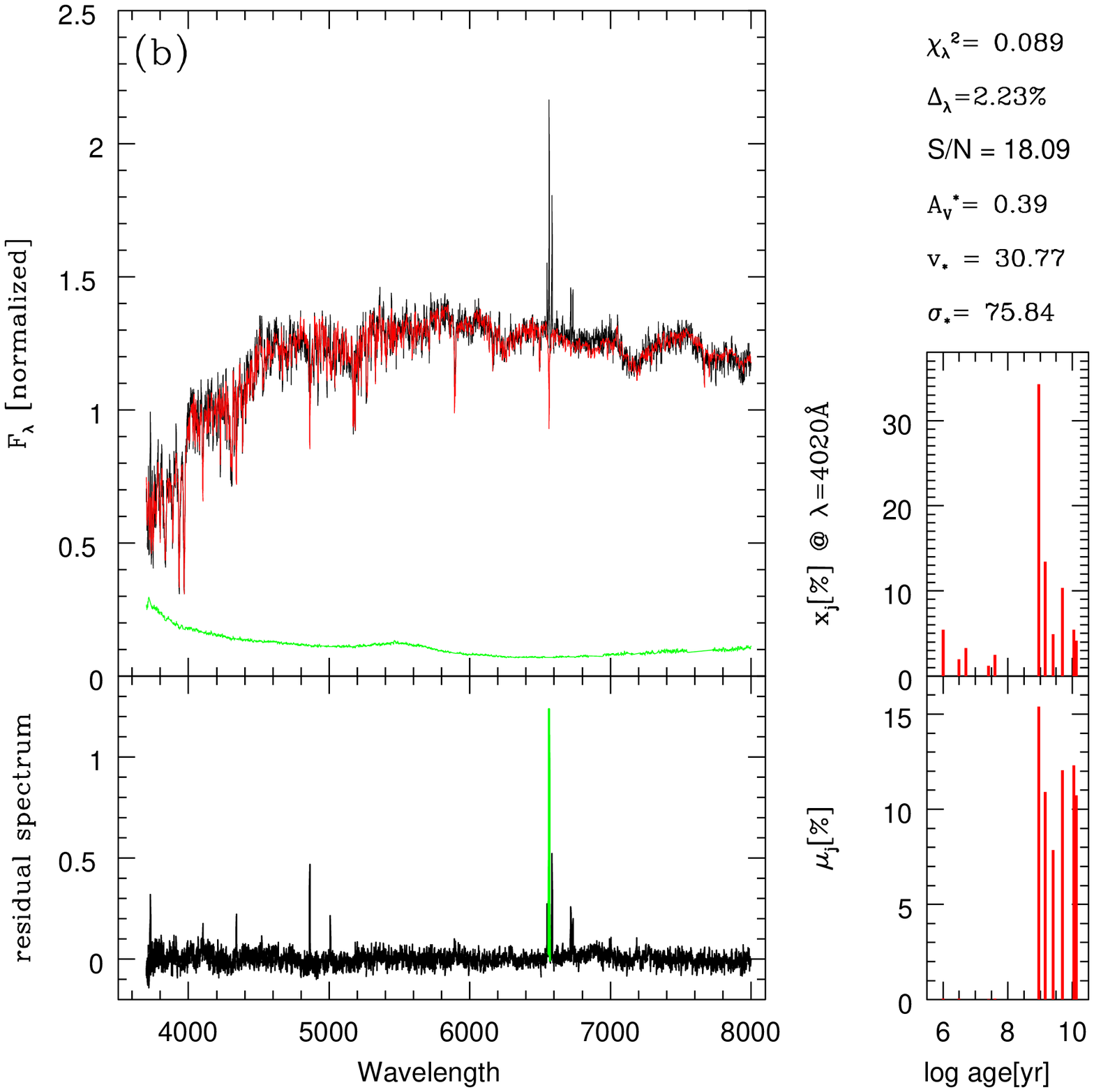}
\includegraphics [width=7.2cm, height=7.0cm] {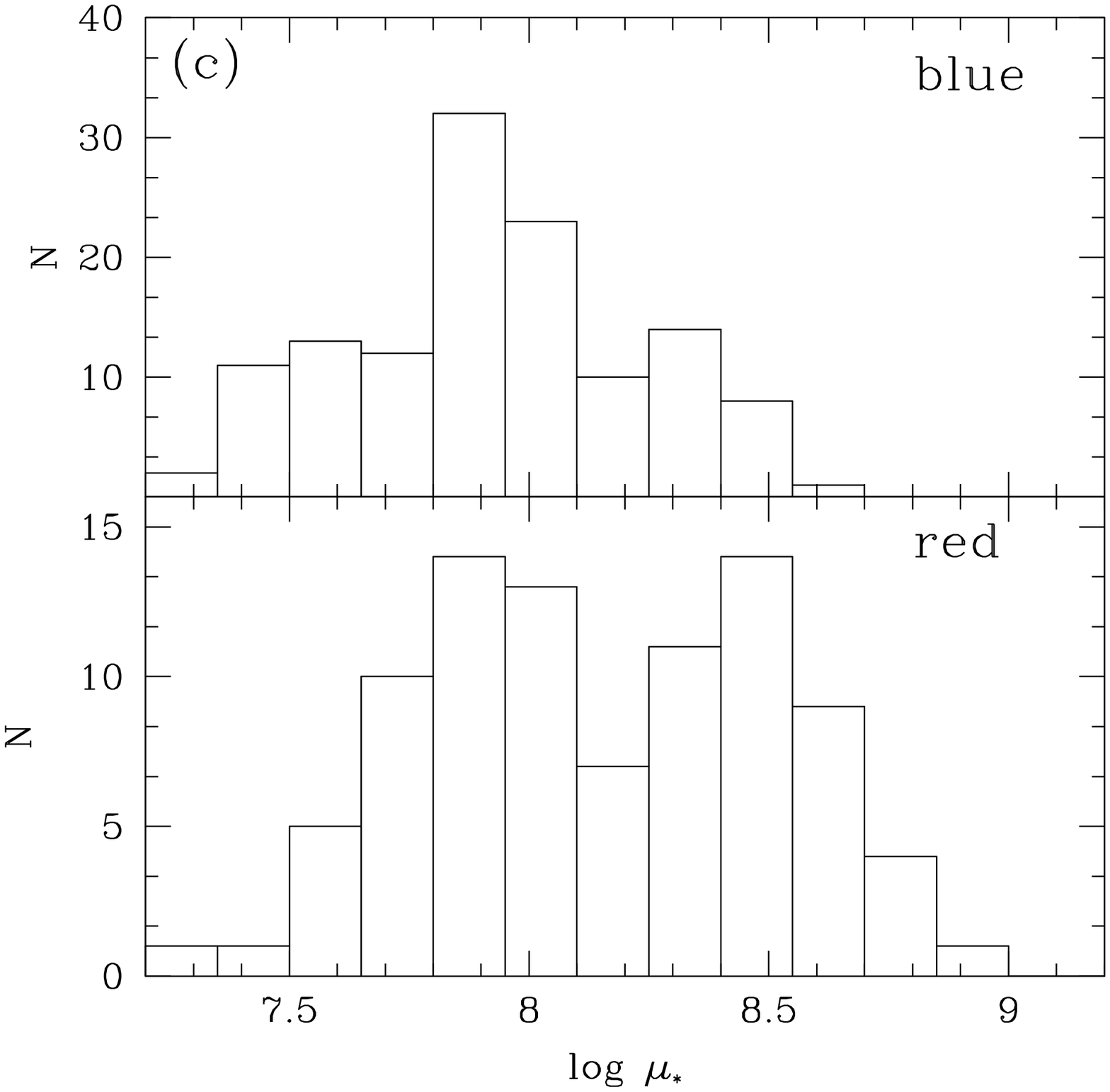}
\includegraphics [width=7.2cm, height=7.0cm] {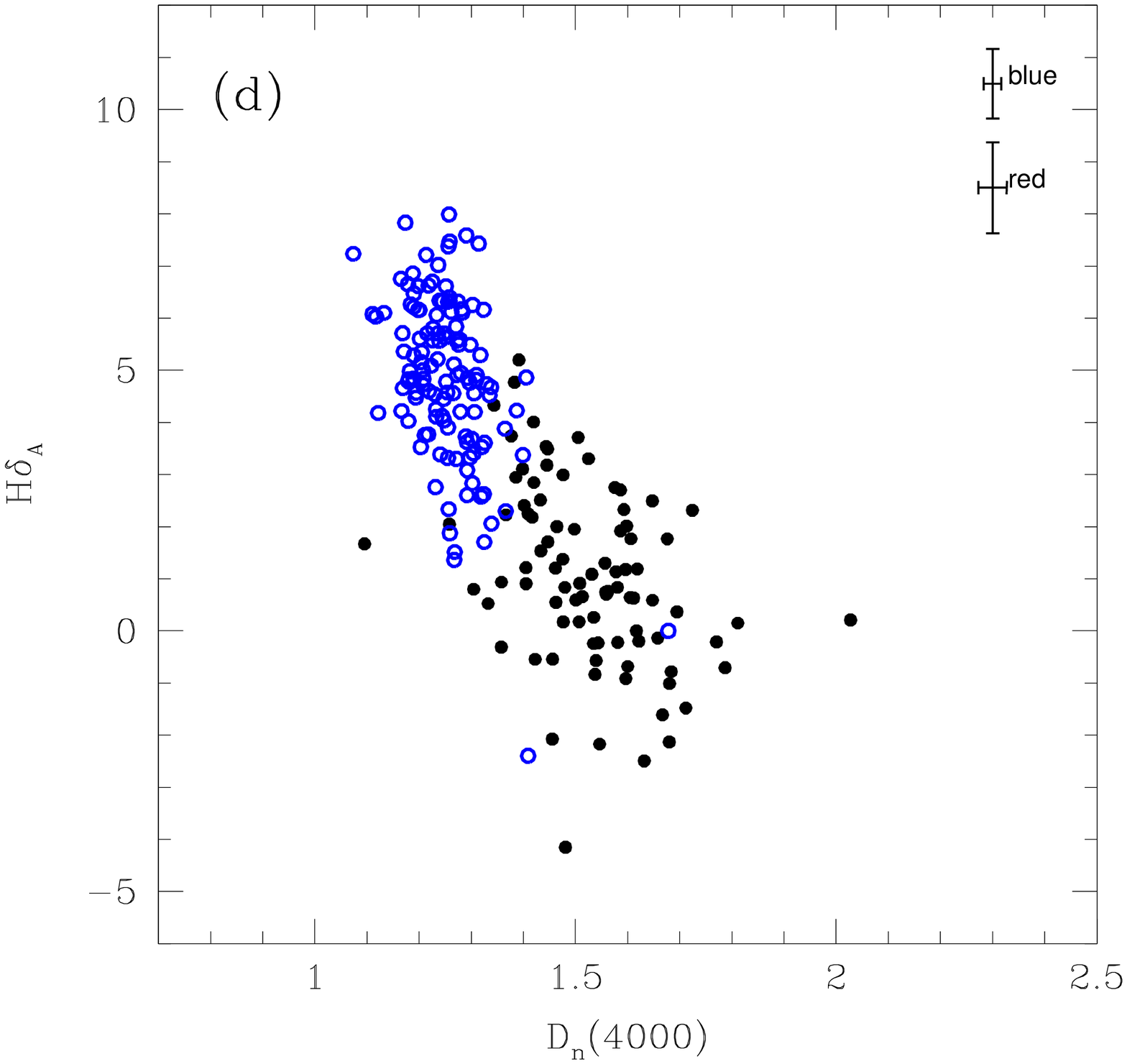}
\caption{The properties of mass limiting sub-samples for red and blue LSBGs: (a). the same as Fig.~\ref{fig.2}, but
for the sub-sample of blue LSBGs;  (b). the same as Fig.~\ref{fig.2} for the sub-sample of blue LSBGs; (c).
histogram distribution of surface mass densities for the sub-samples of red (bottom) and blue (top)
LSBGs; (d). the relations between $D_n$(4000) and stellar mass for sub-samples.}
\label{fig.4}
\end{center}
\end{figure*}

In order to be a more fair comparison between red and blue LSBGs, we select two mass limiting sub-samples of
red and blue LSBGs from our total sample (rather than the surface brightness limiting sub-sample with $\mu_0(R)$, 
since if we select mass limiting sub-samples from the surface brightness limiting sub-sample, the sources of 
red LSBGs will be very small. Considering the overlap of stellar mass in Fig. \ref{fig.1}b and Fig. \ref{fig.1}c 
between red and blue LSBGs, we define the range of 9.5 $\leq$ log$(M_\star/M_\odot)$ $\leq$ 10.3 for both red 
and blue LSBGs. Then we obtain 83 red and 120 blue LSBGs in this stellar mass range as our two mass limiting 
sub-samples for more comparisons.

In Fig. \ref{fig.4}, we show the properties of the two sub-samples of red and blue LSBGs. Fig. \ref{fig.4}a 
and Fig. \ref{fig.4}b are the same as Fig. \ref{fig.2}a and Fig. \ref{fig.2}b, but for the two mass limiting 
sub-samples. 
Table \ref{table2} (right part) presents the contributed light fractions of the stellar populations 
to the blue and red LSBGs. They are very similar to those of the total sample
as given in Table \ref{table1}, less than 3\% discrepancy in the stellar populations
with age bins.
Fig. \ref{fig.4}c shows the histogram distribution of surface mass densities for our mass limiting sub-sample.
Red LSBGs also have a little higher surface mass densities than blue LSBGs, with median values of 1.6 $\times$
10$^8$ and 7.4 $\times$ 10$^7$, respectively. 
These are also similar to the total sample,
and to the surface brightness limiting sub-sample as well.
Moreover, this mass limiting sub-samples also do not show much difference in the distribution of the $D_n$(4000)-$H\delta_A$
plane, just with less scatter  (Fig. \ref{fig.4}d). 
This also confirms that the SFH properties are different between red and blue LSBGs as Fig. \ref{fig.2}d showed.

\section{Discussions}
\label{sec.6}

We discuss the uncertainties of spectral synthesis and the aperture effects in this section.

\subsection{Uncertainties of Spectral Synthesis}

 We discuss the uncertainties of the stellar populations of galaxies
calculated by using STARLIGHT code and the SSP templets from Bruzual \& Charlot (2003).
        In our earlier work, Chen et al. (2010),
	 by using STARLIGHT to compare six popular evolutionary stellar population 
	 synthesis models, we have also discussed the uncertainties of the resulted
	 stellar populations.

 	 As we know, STARLIGHT group has used this code to analyze
	 large sample of SDSS galaxies as shown in their series of work
	 (Cid Fernandes et al. 2005, 2007; Mateus et al. 2006; 
	 Asari et al. 2007). And they have tested the uncertainties 
	 of the resulted stellar populations.
	 In the study of Cid Fernandes et al. (2005), they found 
	 uncertainties are smaller than 0.05, 0.1, and 0.1 for 
	 young ($t < 10^8$), intermediate (10$^8 < t < 10^9$), 
	 and old ($t > 10^9$) populations for S/N$\geq$10, respectively.
	 In our fittings on the sample galaxies, the code provides the values 
	 (e.g. the contributed light fractions of SSPs) of 
	 last-chain-values for 7 Markov chains, and we find that most of
	 the discrepancies in these adopted values are less than 1\%.
	 Thus we believe the uncertainties of the resulted light fractions 
	 are small, and we consider the uncertainties of the resulted 
	 stellar populations as insignificant.
	      
In the fittings,
most of the age sensitivity comes from the continuum shape and $D_n$(4000) break, thus
degeneracy with dust should be discussed carefully.
The dust extinction has been considered in STARLIGHT
	 as a variable parameter, and we adopt the extinction law of 
	 Cardelli et al. (1989) for the code. 
	 The resulted dust extinction $A_V$ from the code is around
	 0.4 for the sample galaxies as given in Fig.~\ref{fig.2} and Fig.~\ref{fig.4}.
	 We have also measured the $A_V$ value from H${\alpha}$/H${\beta}$ for the blue LSBGs from 
	 the combined spectra as given in Fig.~\ref{fig.2}a. 
	 It is just about 0.4, consistent with
	 the resulted one from STARLIGHT program. Thus, the dust
	 extinction effect could have been reliably considered in the
	 the stellar population analyses here. 
	 
       Furthermore, we use a simpler method 
	 to test the age and metallicity degeneracy in the resulted stellar populations
	 from STARLIGHT. 
	 We use 15 SSPs with 15 ages at $Z = Z_\odot$ to do the spectral synthesis on the blue and red LSBGs,
	 and then use another 15 SSPs with same ages but at $Z = 0.2 Z_\odot$
	 to re-do the spectral synthesis. 
	 Comparing these results with those from all three $Z$ case (Table 1), for blue LSBGs, 
	 the $Z = Z_\odot$ results show $\sim$12\% more young population, 
	 and $\sim$7\% less intermediate population;
	 and the $Z = 0.2 Z_\odot$ results show $\sim$9\% less young population,
	 and  $\sim$10\% more old population; for red LSBGs, the light fractions have little changes.
	 Thus, from these analysis the metallicity will cause 
	 uncertainties about 10\% for the stellar populations of blue LSBGs,
	 and very small effect on red LSBGs.
	 But here we prefer the 3 metallicity results.
	 In any case, the dominant population of the sample galaxies are unchanged,
	 i.e. the young population.

\subsection{Aperture effects}

\begin{figure}
\begin{center}
\includegraphics [width=8.0cm, height=7.2cm] {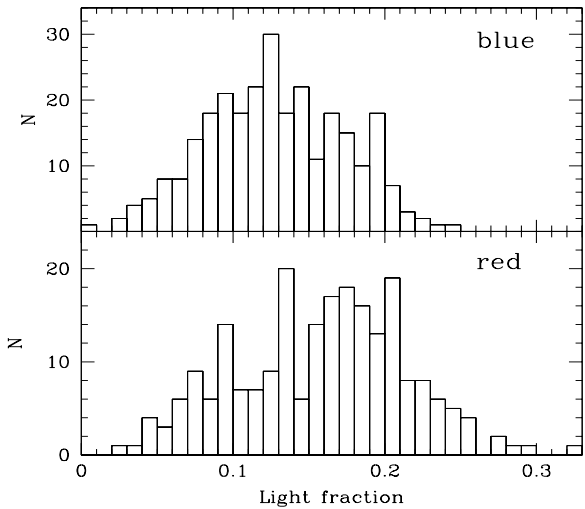}
\caption{The histogram distributions of the light fraction (Eq. \ref{eq.fiber}) with bins of 0.01 for both
red (bottom) and blue (top) LSBGs of our total sample. The median values of the light fractions are 0.16 for red LSBGs and 
0.13 for blue LSBGs.}
\label{fig.5}
\end{center}
\end{figure}

 The SDSS is a fiber based survey. Therefore, we consider the aperture effects in this section. 
Tremonti et al. (2004) and Kewley et al. (2005) have discussed the weak effect of 
the 3{$^{\prime \prime }$}  aperture of
SDSS spectroscopy. They believe that redshifts $z >$ 0.03 and 0.04 are required for SDSS galaxies, respectively
to get reliable metallicities. 
Given that the fiber mag is a measurement of the light going down the fiber and the petrosian mag is a 
good estimate of the total magnitude. Thus, to check how much the light of the galaxies was covered by the 
fiber observation, one simple and accurate way is to compare the ``fiber" and ``petrosian" magnitudes of the 
SDSS galaxies (Liang et al. 2010). We choose the formula:
\begin{equation}
\label{eq.fiber} 
light\_fraction = 10^{(-0.4\times(fiber\_mag - petro\_mag)_r)},
\end{equation}
to estimate how much light was covered by the fiber observations. Fig. \ref{fig.5} shows the light fractions for 
red and blue LSBGs of our total sample. It shows that the light fractions of red (bottom) and blue (top) LSBGs are 0.16 and 0.13,
respectively. Therefore, the properties of red and blue LSBGs are main from the central regions, which may be redder,
older and more metal-rich than the outer regions. However, it could not affect 
much our results of the differences between
red and blue LSBGs, because the light fractions are almost same.
Moreover, all our sample galaxies are disk-dominated galaxies
	 with very small bulges, we assume the resulted stellar 
	 populations could be good representatives of their disk populations. 

\section{Summary}
\label{sec.7}

Our main results can be summarized as follows. We present a large sample of 213 red and 266 blue disk LSBGs 
from SDSS-DR7, which have sufficient S/N in the spectral continua to study their SFH by using spectral synthesis 
through STARLIGHT code and the SSPs of Bruzual \& Charlot (2003), as well as the absorption-line indices ($Mg_2$,
$H\delta_A$) and $D_n$(4000). 

\begin{enumerate}

\item Blue LSBGs are dominated by young populations with a few old populations (5.9\% populations
older than 5 Gyr), however, there are a significant fraction of old populations in red LSBGs (33.3\%). 
The dominated populations of blue LSBGs are $Z = 0.2Z_\odot$, while the dominated populations of 
red LSBGs are $Z = Z_\odot$, and red LSBGs are more metal-rich.
Red LSBGs tend to be more massive and have higher surface mass density than blue LSBGs.

\item The $D_n$(4000)-$H\delta_A$ plane shows that red LSBGs have different SFH from blue LSBGs: blue
LSBGs are more likely to be experiencing a sporadic star formation events at the present day, whereas
red LSBGs are more likely to form stars continuously. Moreover, the fraction of galaxies that
experienced recent sporadic star formation events deceases with increasing stellar mass.

\item By defining two sub-samples according to surface brightness $\mu_0(R)$
and stellar mass limits for both blue and red LSBGs, i.e.
the $\mu$-sample with $\mu_0(R)$ $\geq$ 20.7 mag arcsec$^{-2}$,
and the M-sample with 
9.5 $\leq$ log$(M_\star/M_\odot)$ $\leq$ 10.3, 
we find that they show very similar results to the total sample (T-sample) on the
$D_n$(4000)-$H\delta_A$ plane, surface mass density
and stellar populations etc. These confirm well that the comparisons for blue and red LSBGs we worked
are robust. 

\end{enumerate}

The Large Synoptic Survey Telescope (LSST, Paul et al. 2009) should be sensitive to galaxies with central
surface brightness as low as 27 mag arcsec$^{-2}$ in the $r$-band in the ten-year stack-compared with SDSS,
where the faintest galaxies measured have central surface brightness $\mu_r$ $\sim$ 24.5 mag arcsec$^{-2}$
(Zhong et al. 2008). Moreover, this aspect will also discover larger numbers of giant LSBGs spirals and tie down
the population of red spiral LSBGs. Therefore it is helpful to study the stellar populations and SFH from the LSST data
sets in the future.

\begin{acknowledgements}
We thank the referee for the valuable comments to help improve this work. 
We thank Dr. James Wicker for helping us to correct the English description in the text.
This work was supported by the NSFC grants 10933001, 10973015, 10673002, and the National Basic
Research Program of China (973 Program) grants 2007CB815404, 2007CB815406,
and No. 2006AA01A120 (863 project).
The STARLIGHT project is supported by the Brazilian agencies CNPq,
 CAPES, and FAPESP and by the France-Brazil CAPES/Cofecub program.
 We thank the useful SDSS database and the MPA/JHU catalogs.
\end{acknowledgements}

\clearpage
\end{document}